\documentclass[runningheads]{llncs}
\usepackage[dvipsnames]{xcolor}
\usepackage{float,graphicx}
\usepackage{mathtools}
\usepackage{amsmath}
\usepackage{caption}
\usepackage{subcaption}
\usepackage{tabularx}
\usepackage{hyperref}
\usepackage{enumitem}
\usepackage{wrapfig}

\usepackage{listings,amsfonts}

\usepackage{tikz}
\usetikzlibrary{arrows,automata}
\usetikzlibrary{positioning}
\tikzset{initial text={},auto} 
\tikzstyle{every state}=[draw,shape=circle,inner sep=1pt,minimum size=8pt]

\newcommand{\ttrue}{\mathtt{tt}}
\newcommand{\ffalse}{\mathtt{ff}}

\usepackage{courier} 

\lstdefinelanguage{mCRL2}
{
keywords={act,var,cons,end,eqn,glob,init,val,whr,sort,map,pbes,proc,struct},
keywords=[2]{true,false,delta,tau},
keywords=[3]{Bool,Nat,Real,Pos,Int,Set,Bag,List,Int2Nat,Pos2Nat,Int2Pos,min,max
},
keywords=[4]{hide,if,rename,sum,in,mu,nu,forall,exists,mod,allow,block,comm},
keywords=[5]{nested,initial},
numberstyle=\color{blue},
comment=[l]\%,
commentstyle=\slshape,
keywordstyle=[1]\bfseries,
keywordstyle=[2]\itshape,
keywordstyle=[3]\itshape,
keywordstyle=[4]\itshape,
keywordstyle=[5]\bfseries\itshape,
basicstyle=\ttfamily\scriptsize,
flexiblecolumns=false,
breaklines=false,
tabsize=2,
literate={\ \ }{{\ }}1,
mathescape=true
}
[keywords,comments]

\lstdefinelanguage{mCRL2-inline}
{
keywords={act,var,cons,end,eqn,glob,init,val,whr,sort,map,pbes,proc,struct},
keywords=[2]{true,false,delta,tau},
keywords=[3]{Bool,Nat,Real,Pos,Int,Set,Bag,List,Int2Nat,Pos2Nat,Int2Pos,min,max
},
keywords=[4]{hide,if,rename,sum,in,mu,nu,forall,exists,mod,allow,block,comm},
keywords=[5]{nested,initial},
numberstyle=\color{blue},
comment=[l]\%,
commentstyle=\slshape,
keywordstyle=[1]\bfseries,
keywordstyle=[2]\itshape,
keywordstyle=[3]\itshape,
keywordstyle=[4]\itshape,
keywordstyle=[5]\bfseries\itshape,
basicstyle=\ttfamily\footnotesize,
flexiblecolumns=false,
breaklines=true,
mathescape=true
}
[keywords,comments]

\lstdefinelanguage{Cordis}
{
keywords={Cmd, Input, InputSignal, Machine, MachinePart, Message, Observer, OutputSignal, Output, Setting, Var},
basicstyle=\ttfamily\footnotesize,
keywordstyle=\ttfamily\footnotesize,
flexiblecolumns=false
breaklines=true,
breakatwhitespace,
mathescape=true
}
[keywords]

\newcommand{\inlinemcrl}{\lstinline[language=mCRL2-inline]} 
\newcommand{\inlinecordis}{\lstinline[language=Cordis]} 
\newcommand{\false}{\ensuremath{\mathit{false}}}      
\newcommand{\sm}[1]{\inlinemcrl{#1}}             
\newcommand{\stereotype}[1]{\inlinecordis{<<#1>>}} 
\newcommand{\true}{\ensuremath{\mathit{true}}}        

\usepackage[obeyFinal]{todonotes}
\usepackage[most]{tcolorbox}
\definecolor{ruddypink}{rgb}{0.88, 0.56, 0.59}

\begin{document}

\title{Formal verification of an industrial UML-like model using mCRL2 (extended version)}

\author{Anna Stramaglia \and
Jeroen J.A. Keiren\orcidID{0000-0002-5772-9527}
}

\institute{Eindhoven University of Technology, the Netherlands\\ \email{\{a.stramaglia, j.j.a.keiren\}@tue.nl}}

\maketitle

\begin{abstract}
Low-code development platforms are gaining popularity. Essentially, such platforms allow to shift from coding to graphical modeling, helping to improve quality and reduce development time. The Cordis SUITE is a low-code development platform that adopts the Unified Modeling Language (UML) to design complex machine-control applications. In this paper we introduce Cordis models and their semantics. To enable formal verification, we define an automatic translation of Cordis models to the process algebraic specification language mCRL2.
As a proof of concept, we describe requirements of the control software of an industrial cylinder model developed by Cordis, and show how these can be verified using model checking. We show that our verification approach is effective to uncover subtle issues in the industrial model and its implementation.
\end{abstract}

\section{Introduction}\label{sec:intro}
Abstract models are commonly used during the design phase of software. For example, class diagrams are used to describe the structure of a software system, and behavioral models describe the possible executions.
Model checking can be used to verify that such a behavioral model satisfies its requirements.
While model checking is a promising technique, its industrial applications are still limited.
There are several reasons for this, of which we name two examples here.
First, it is considered tedious to create a detailed behavioral model prior to implementing the system.
Second, the available model checking tools primarily use low-level, academic languages that require specific expertise not typically acquired by engineers in industry.

Low-code development platforms (LCDPs)~\cite{sahay_supporting_2020} are gaining popularity.
Such platforms focus on increasing the level of abstraction of software development, shifting from coding to graphical modeling.
Executable code is typically generated from low-code models.
LCDPs allow us to address both issues described above.
First of all, the detailed behavioral model is now created during implementation of the the system: the model is the artifact from which the executable code is automatically generated.
Second, provided that the semantics of the models are well-understood, the models are sufficiently detailed that they can be automatically translated to the input languages used by state-of-the-art model checkers.

The Cordis SUITE\footnote{https://www.cordis-suite.com} is a complete LCDP for machine-control applications, based on graphic Model-Driven Software Engineering, describing system structure and behavior in UML diagrams~\cite{UML2.5.1}.
The development environment in the Cordis SUITE is the Cordis Modeler, which uses Altova UModel\footnote{https://www.altova.com} as a front-end for drawing the models.
Cordis models use extensions of UML class diagrams for describing the static structure and UML state machine diagrams for the behavior.
The features used in Cordis models are fairly rich: besides the standard UML models, it comprises several Cordis specific extensions and a large fragment of the Structured Text language~\cite{john_programming_2010}.
Ultimately, using the Cordis SUITE, it is possible to generate source code for Programmable Logic Controllers (PLCs) or the .NET platform directly from the Cordis models.
Hence, the implementation of Cordis models is consistent, in terms of execution, with their design.
Cordis, the company developing this LCDP, has shown an interest in extending the Cordis SUITE with model checking capabilities.

Our contributions in this paper are as follows.
To enable model checking of Cordis models, we first describe their structure and semantics.
We describe an automated translation from Cordis models to the mCRL2 specification language~\cite{groote2014modeling}.
The use of mCRL2 is motivated by the availability of its associated tool set with powerful verification tools such as simulation, explicit model checking and the verification of modal $\mu$-calculus formulae~\cite{mCRL2}.
For verification, we use a first-order extension of the modal $\mu$-calculus~\cite{GM1999}.
We illustrate the feasibility of modeling and verification of Cordis models using a pneumatic \emph{cylinder}.
We specify, both informally and formally, two typical requirements of the cylinder, and verify whether its model satisfies these requirements.
One of the requirements is not satisfied by the initial model.
Using mCRL2 and its powerful counterexamples, we analyze why this requirement is not satisfied.
The analysis uncovers and confirms a subtle issue in the cylinder model, that can, indeed, be reproduced in the PLC code of the implementation. We propose a fix that has been integrated into the model distributed by Cordis.

\paragraph{Related work.}
A large amount of work has been done in the application of formal verification to industrial domains.
Most of this work focuses on specific domains, such as railway infrastructure management~\cite{bouwman_formalisation_2021,hansen_towards_2010,salunkhe_automatic_2021} and medical applications~\cite{KK12,pore2021safe,santone_radiomic_2021}.
More closely related to this paper are works on the modeling and verification of control software.
For instance, CERN's FSM language~\cite{HKKLW13} uses a strict hierarchical architecture of finite state machines tailored towards a specific machine control application.
The OIL language, developed and used by Canon Production Printing, has a strong focus on separation of concerns during the development~\cite{bunte_formal_2020}.

Modeling languages such as SysML and UML can be used to model systems that come from any domain.
The verification of these languages, in particular their state machine diagrams, has been studied extensively, see for example~\cite{bouwman_formalisation_2021,dubrovin_symbolic_2008,hutchison_formal_2013,lyazidi_formal_2019,giannakopoulou_execution_2014,santos_transformation_2014,schafer_model_2001}.
Of these, the work by Bouwman et al.~\cite{bouwman_formalisation_2021} is closest to ours.
Their focus is on the verification of SysML state machines in the railway industrial domain.
Like in our work, they describe a formal semantics for state machines, and implement a translation to mCRL2 to allow formal verification. The semantics and execution model in that work focuses on distributed execution of state machines that communicate via queues, whereas our work focuses on a strictly sequential execution where communication takes place via shared variables.
Lyazidi and Mouline in~\cite{lyazidi_formal_2019}, define a transformation from UML state machine diagrams to Petri nets.
Santos et al. in~\cite{santos_transformation_2014}, present an approach to transform various UML behavioral diagrams into a single transition system for the model checker NuSMV~\cite{goos_nusmv_2002}, in order to support model checking.

\paragraph{Outline.}
In Section~\ref{sec:cordis-models} we explain the structure and semantics of Cordis models. In Section~\ref{sec:cylinder} we present the cylinder model.
In Section~\ref{sec:mcrl2translation_verification} we describe the mCRL2 specification of Cordis models, and the requirements verified against the cylinder model, and the results obtained from the verification of the cylinder model. Discussion and conclusions are presented in Section~\ref{sec:discussion} and Section~\ref{sec:conclusion}, respectively.

\section{Cordis models}\label{sec:cordis-models}
Cordis models are developed, tested and simulated in the Cordis SUITE.
In this section we give a brief introduction of the Cordis SUITE and subsequently we describe the structure and semantics of Cordis models.

\subsection{Cordis SUITE}\label{overviewSUITE}
The Cordis SUITE is a collection of integrated tools for developing, testing and deploying system control software, with a particular focus on machine control.
The main components and their connections are shown in Figure~\ref{fig:Suite}.

\noindent
\begin{minipage}{0.52\textwidth}
\smallskip
The \textit{Cordis Modeler} is an LCDP for creating machine-control applications. It uses an extension of the Unified Modelling Language (UML) such that every Cordis model describes the structure and behavior of a \emph{machine}. In particular, the modeler uses class diagrams for the static structure, and state machine diagrams for the behavior. Additionally, it can check for design errors, and generate source code for Programmable Logic Controllers (PLCs) or the .NET platform.
\end{minipage}\hfill
\begin{minipage}{0.48\textwidth}
  \begin{center}
    \includegraphics[width=0.8\textwidth]{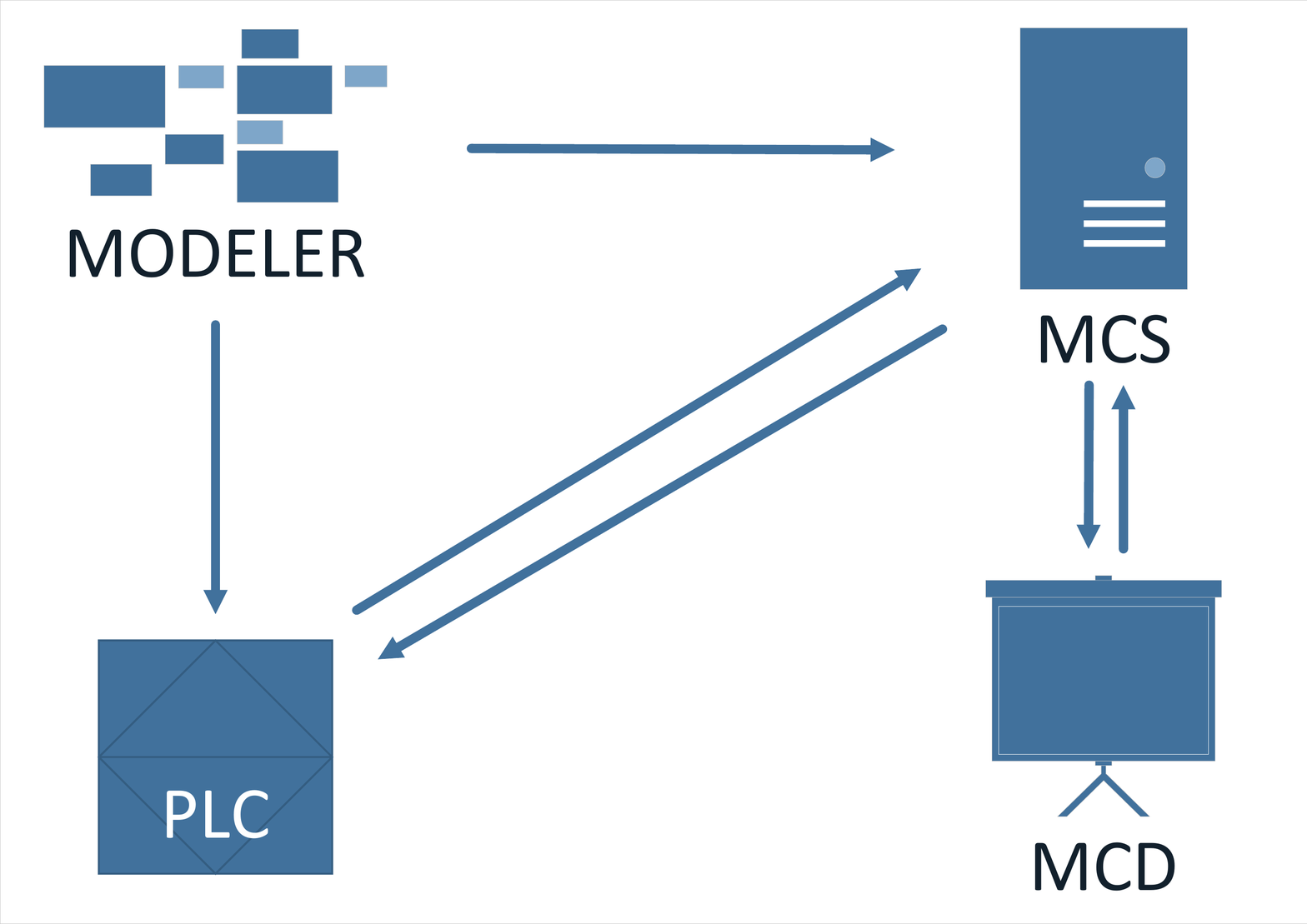}
  \end{center}
  \captionof{figure}{Main components of Cordis \\SUITE and their connections.}
  \label{fig:Suite}
\end{minipage}
\noindent
The generated code can subsequently be compiled and executed on the target platform.
The \textit{Cordis Machine Control Server (MCS)} loads model information from the modeler, and connects to the PLC in order to exchange state information and data with the running system.
The \textit{Cordis Machine Control Dashboard (MCD)} is a Human-Machine Interface used to monitor the live system data and diagrams made available through the MCS, and to send inputs to the system.
In particular, the MCD shows live state machine diagrams when the PLC is running, providing the user with real-time and historical information about the execution of a state machine.

The Cordis SUITE's core element is the Cordis Modeler, and the structure and semantics of Cordis models is the focus of the following subsections.

\subsection{Class diagrams}
The \emph{static structure} in a Cordis model is described using a class diagram.
Classes can be tagged with stereotypes \stereotype{Machine}, denoting the machine that is controlled by the system, and \stereotype{MachinePart}, denoting components of the system.
A class has \emph{properties} and \emph{operations}.
Class properties are the variables stored in the class and are tagged with Cordis-specific stereotypes that describe their role in the system.
Stereotypes \stereotype{InputSignal} and \stereotype{OutputSignal} are used to define shared variables that are used to communicate between objects within the model; \stereotype{Input} and \stereotype{Output} describe variables that are used to interface with the environment, typically the hardware.
Variables that are only used from within the object have stereotype \stereotype{Var}.
Properties of type \stereotype{Setting} can be configured from the MCD; these are essentially configuration parameters of the system that are constant throughout the standard execution.
The system can provide feedback to the user using \stereotype{Message} properties.
Only properties with stereotype \stereotype{Observer}, \stereotype{Setting} and \stereotype{Message} are visible in the MCD.
All class operations are commands which are issued (asynchronously) to the class, either by the environment or another component in the system. This is described by stereotype \stereotype{Cmd}.
\begin{figure}
\centering
\includegraphics[width=0.6\textwidth]{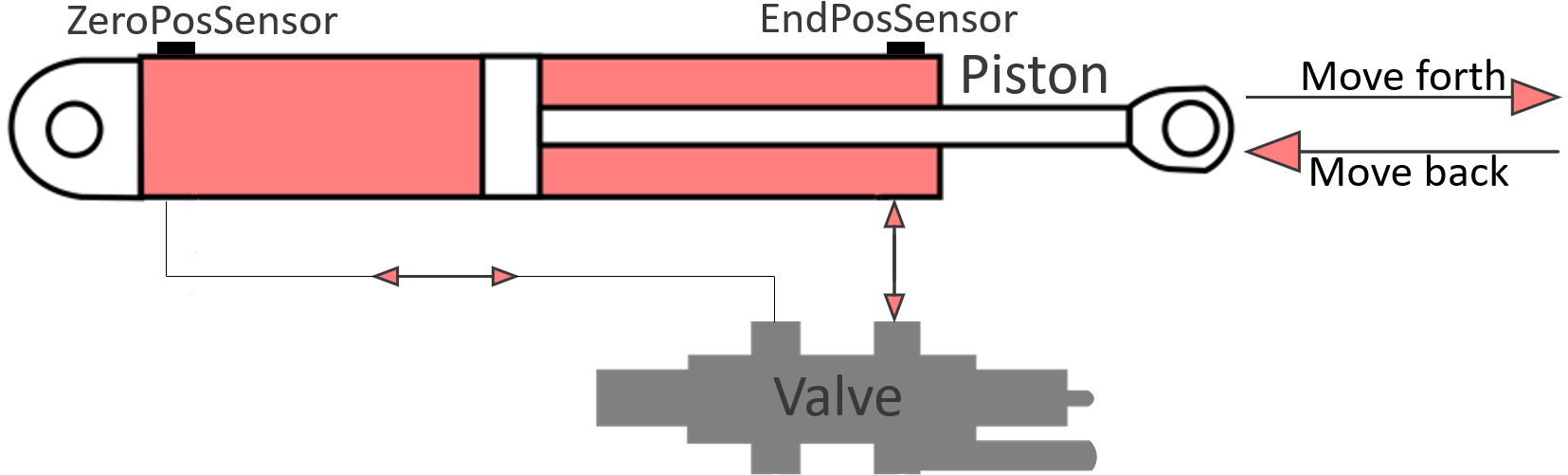}
\caption{Pneumatic cylinder~\cite{neves_hydraulic_2011}}
\label{fig:cylinder}
\end{figure}
\begin{example}\label{ex:cylinder-class}
In this paper we consider the Cordis model of a pneumatic cylinder to illustrate the syntax and semantics of Cordis models, as well as their verification.
Pneumatic cylinders are commonly used in factory automation systems for mechanisms such as clamping, ejecting, blocking, and lifting, and in industrial processes for materials handling and packaging.
A pneumatic cylinder, see Figure~\ref{fig:cylinder}, consists of a cylinder barrel, with a piston which moves back or forth. It is powered using compressed air, whose flow is controlled by electrically controlled valves.
The cylinder we consider in this paper moves the piston between the \emph{zero position} and the \emph{end position}, in which it is completely retracted or extended, respectively. It is equipped with sensors, here \emph{ZeroPosSensor} and \emph{EndPosSensor}, to detect the current position of the piston.

The cylinder is typically used as a machine part in a larger machine. For the sake of simplicity, in this paper we consider the cylinder in isolation. Due to its simplicity, its (trivial) class diagram consists of a single class, shown in Figure~\ref{fig:cylinder-class}.

\begin{figure}[ht]
\centering
\includegraphics[width=0.45\textwidth]{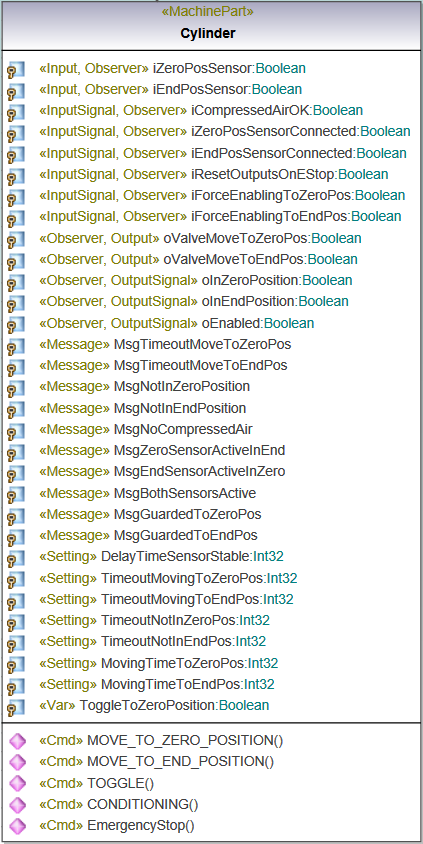}
\caption{\inlinecordis{Cylinder class}}
\label{fig:cylinder-class}
\end{figure}

The cylinder has two inputs, \inlinecordis{iZeroPosSensor} and \inlinecordis{iEndPosSensor} that detect whether the cylinder is at its zero or end position, respectively. It has two outputs, \inlinecordis{oValveMoveToZeroPos} and \inlinecordis{oValveMoveToEndPos} that are used to actuate the valves.
Furthermore, it has a number of input signals, i.e., shared variables that are written by other objects in the machine that uses the cylinder and that are read by the cylinder, and output signals, i.e., shared variables that are written by the cylinder, and used by other objects. Property \inlinecordis{ToggleToZeroPosition} is used internally to toggle the direction of movement of the cylinder.
The system also has a number of settings and messages that are irrelevant to the verification. We will ignore these in the rest of this paper.

The cylinder has five commands that can be used to control the cylinder.
The meaning of \inlinecordis{MOVE_TO_ZERO_POSITION}, \inlinecordis{MOVE_TO_END_POSITION}, \inlinecordis{TOGGLE}, and \inlinecordis{EmergencyStop} are self-explanatory. Command \inlinecordis{CONDITIONING} can be used to force (re)initialization of the cylinder.
\end{example}

\subsection{State machine diagrams}\label{sec:state-machines}
The behavior of an object is defined using a hierarchy of state machine diagrams.
We first describe the structure of such diagrams, and subsequently explain their semantics.

\subsubsection{State machine diagram structure}
Structurally, Cordis state machine diagrams are similar to those defined in standard UML~\cite{UML2.5.1}, but there are some Cordis specific details.

At the highest level, a state machine diagram consists of \textit{top-level state machines}.
Each state machine diagram requires at least one top-level state machine.
A (top-level) state machine consists of a hierarchy of states and pseudo-states, connected by transitions. States and pseudo-states capture the most commonly used (pseudo-)state types described in the UML standard, e.g., simple states, composite states, initial pseudo-states, final states and choice nodes. A state can contain entry and continuous behavior. A state can also be a reference to a \emph{subdiagram}, whose content can, essentially, be syntactically substituted in the diagram that refers to it.
If a state is a reference to a \emph{substatemachine}, the substatemachine that is referenced is executed separately from the statemachine that references it; when a transition to a substatemachine is taken, control is transferred to the substatemachine. So, the key difference between subdiagrams and substatemachines is whether they are executed as part of the diagram that references it (subdiagram) or not (substatemachine).

Transitions have a \emph{source} and \emph{target} state, and can optionally be labelled by \emph{guards} and \emph{actions}.
A guard is a Boolean condition that must be satisfied for the condition to be enabled. An action typically represents a side effect that is executed upon taking the transition, e.g., updating the value of a class property.
For a transition without guard, the guard is assumed to be \true\ if the source of the transition is an initial state or an exit node, or the target of the transition is a choice node. The guard of an unguarded transition whose source is a choice node is treated as \emph{else}, i.e., it is \true\ only when the guards of all other transitions of the same choice node are \false.
For all other transitions without guard, the guard is assumed to be \false.

\begin{figure}[ht]
\centering
\includegraphics[width=0.95\linewidth]{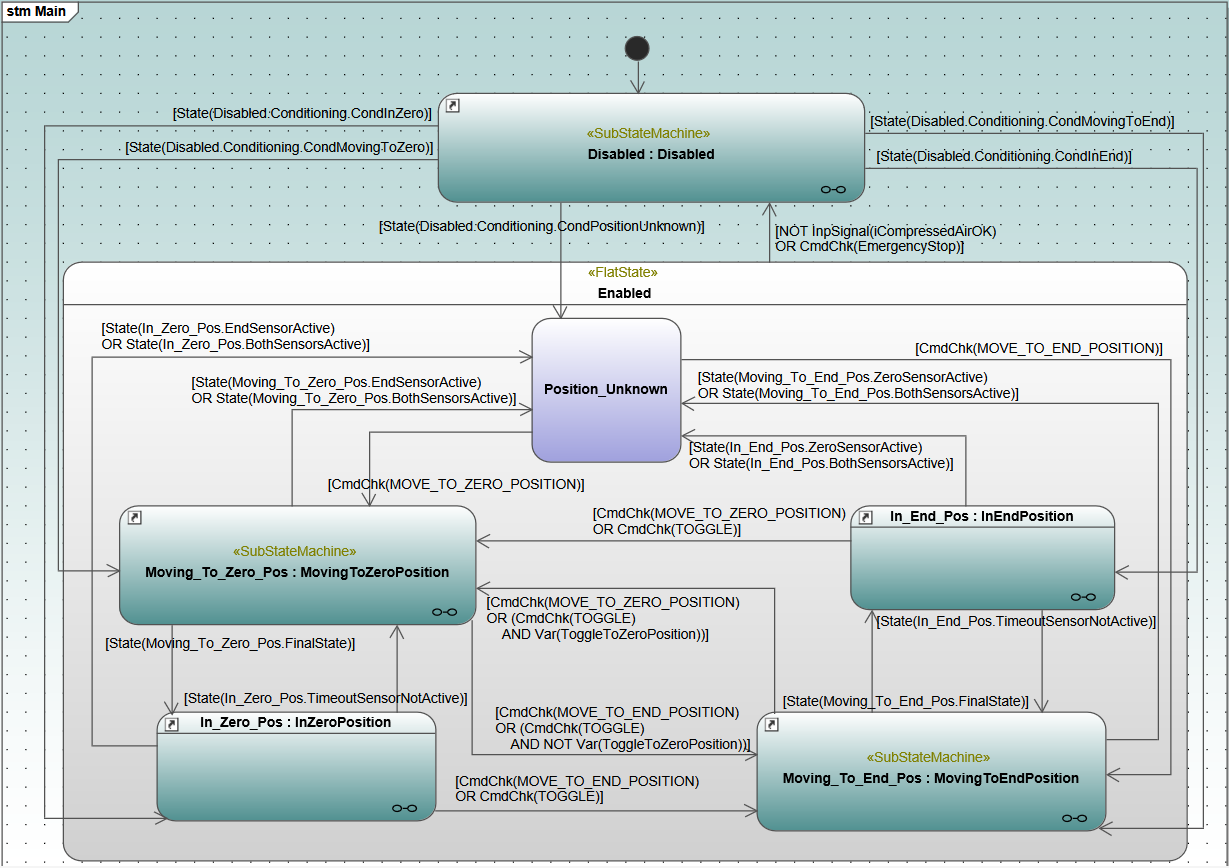}
\caption{State machine \sm{Main}.}
\label{fig:cylinder-main-sm}
\end{figure}

\begin{example}\label{exa:cylinder-main-sm}
Consider the top-level statemachine \sm{Main} of the cylinder from our running example, shown in Figure~\ref{fig:cylinder-main-sm}. Its initial state is \inlinecordis{Disabled}, indicated using the initial pseudo-state $\bullet$. State \inlinecordis{Disabled} is a reference to a substatemachine. This substatemachine has a number of states that are used to model different ways out of the substatemachine (see Figure~\ref{fig:Cond} for details), for instance, if the substatemachine determines the cylinder is in its zero position it reaches state \inlinecordis{CondInZero}, and in state machine \sm{Main} the transition to \inlinecordis{In_Zero_Pos} is taken.

Most transitions have guards, e.g.,  \inlinecordis{Moving_To_Zero_Pos} has a transition to \inlinecordis{In_Zero_Pos} with guard \inlinecordis{[State(Moving_To_Zero_Pos.FinalState)]} which only evaluates to true if substatemachine \inlinecordis{Moving_To_Zero_Pos} is currently in its final state.
Similarly, the guard \inlinecordis{[NOT InpSignal(iCompressedAirOk) OR CmdChk(EmergencyStop)]} is only satisfied if \inlinecordis{iCompressedAirOk} is false, or command \inlinecordis{EmergencyStop} has been accepted by the cylinder.
\end{example}

Some behavior must be executed every time an object is allowed to execute a step, regardless of its current states. This is modeled using pre- and poststates, describing behavior that is executed before and after anything else in the object, respectively. Pre- and poststates can either appear inside a state machine, or at the top-level of the state machine diagram. A state machine diagram can have more than one pre- and poststate, but the behavior of multiple pre- and poststates is always combined into a single prestate and a single poststate by taking the sequential composition of all pre- and poststates in a predefined order.

\begin{figure}
\centering
\includegraphics[width =.5\linewidth]{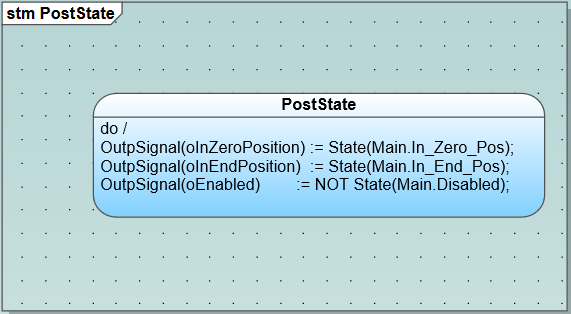}
\caption{\inlinecordis{Poststate} of the cylinder}
\label{fig:post}
\end{figure}

\begin{example}
The cylinder model has a \inlinecordis{PostState} that is shown in Figure~\ref{fig:post}.
It contains the code that is executed after all state machines in the cylinder have executed one step.
It updates the values of output signals \inlinecordis{oInZeroPosition}, \inlinecordis{oInEndPosition} and \inlinemcrl{oEnabled} to reflect the current configuration of the cylinder.
\end{example}

\subsubsection{State machine execution}\label{execution}
Cordis models are executed using a cyclic execution model. In every cycle all objects (i.e., machines and machine parts) take turns executing in a predetermined order, following the hierarchical structure defined in the class diagram.

\begin{figure}[ht]
\centering
\begin{subfigure}{.50\textwidth}
  \centering
  \includegraphics[page=1,width=0.8\linewidth]{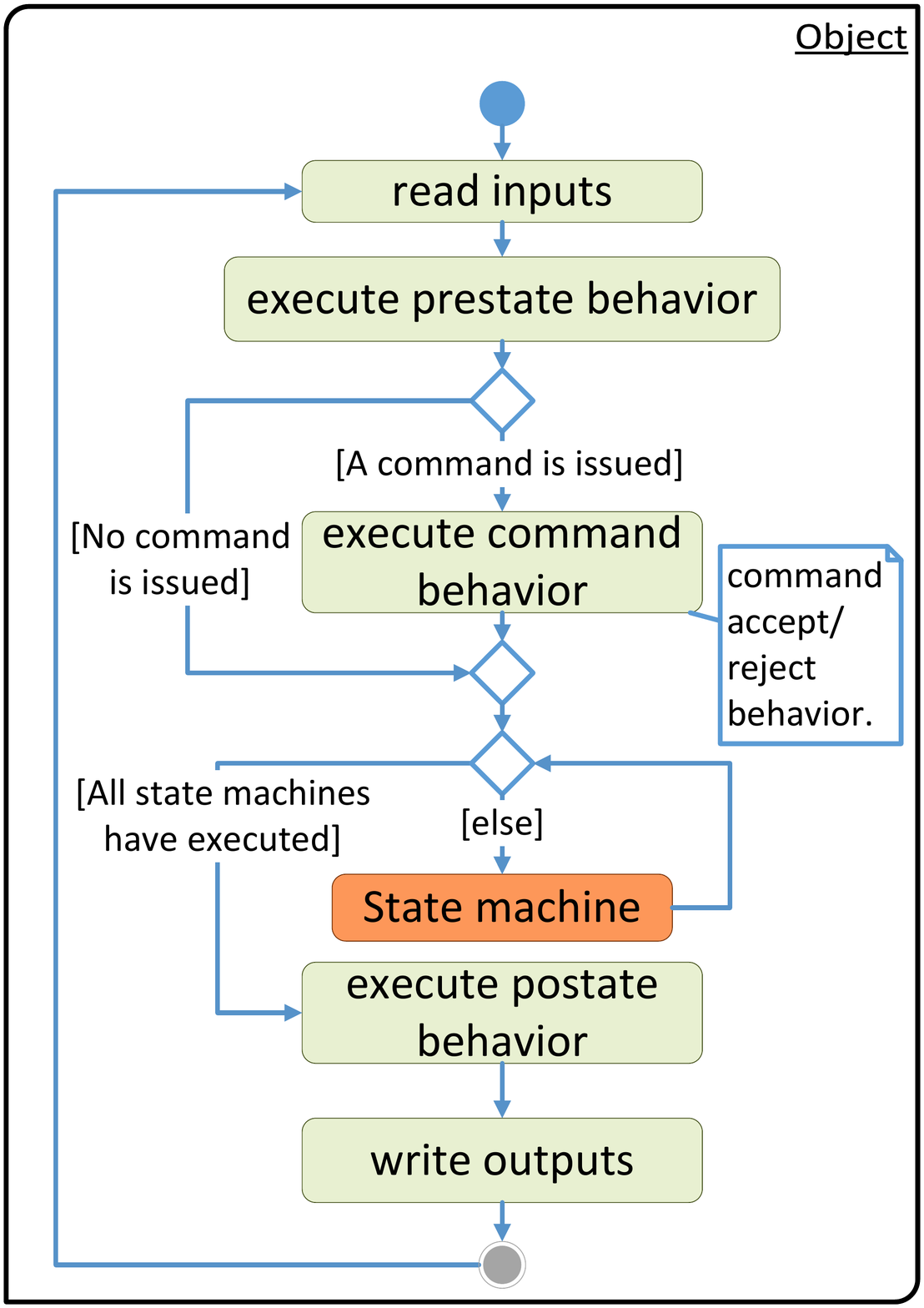}
    \caption{}
  \label{fig:Obj}
\end{subfigure}%
\begin{subfigure}{.50\textwidth}
  \centering
  \includegraphics[page=2,width=0.75\linewidth]{FINALActivityDiagram_CylinderPaper.pdf}
  \caption{}
  \label{fig:Stm}
\end{subfigure}
\caption{Order of execution of (a) one object, and (b) a state machine within the object}
\label{fig:execution-order}
\end{figure}

The order of execution within an object is depicted as an activity diagram in Figure~\ref{fig:Obj}. First the \emph{inputs variables} are read. This essentially caches the current values of the inputs, input signals and commands in local variables. Second, the (combined) prestate of the object is executed.
If a new command was sent to the object, the behavior associated to the command is executed. The execution of a command consists of the following components.
The \emph{guard condition} is a Boolean expression that determines whether the command can be accepted. If the guard condition is true, the \emph{command action} is executed, otherwise the \emph{reject action} is executed.
The \emph{command ready condition} determines whether, at the end of the current cycle, the command has been fully processed and can be removed from the interface of the command. If the command ready condition is false, the command will remain on the interface; if it is not subsequently overwritten by a new command, in the next cycle only the command ready condition will be reevaluated, the guard condition and the accept and reject actions are not executed again.
After the command has executed, all state machines in the object execute in a predetermined order and are allowed to execute one single transition per turn.
Finally, the (combined) poststate is executed and the outputs are written.

Figure~\ref{fig:Stm} depicts the execution of a single state machine using an activity diagram.
First, the continuous code of the current active state is executed. Second, it is determined whether a transition is enabled in the current state of the state machine. A transition is enabled if its guard is true.
If multiple transitions are enabled, only one is executed.
Choosing the transition to be executed is done as follows: if the source state of one enabled transition contains the source state of another, the transition from the outermost state is executed. Transitions to other states take priority over self-loop transitions to the same state.
If, after making this selection, still multiple transitions are enabled, the first transition from a predetermined order is executed.\footnote{Currently, the implementation chooses the order of creation.}
Finally, the current state is changed to the target state and the behavior specified on the transition and the behavior of the target state are executed.
If no transition is enabled, the current state remains unchanged.

For the sake of verification, it is useful to consider the behavior of a subsystem in isolation.
To facilitate this, in addition to the inputs, we also consider the input (and input/output) signals and commands as free variables.
This allows setting arbitrary values to these inputs and commands in the model.

\section{Cylinder}\label{sec:cylinder}
In the remainder of this paper we focus on the verification of the full model of the pneumatic cylinder introduced as running example in Section~\ref{sec:cordis-models}.
Its class diagram and top-level state machine \inlinecordis{Main} were introduced in Examples~\ref{ex:cylinder-class} and~\ref{exa:cylinder-main-sm}, respectively.
In this section, we introduce the remainder of the state machine diagram of the cylinder.

State machine \inlinecordis{Main} refers to substatemachines  \inlinecordis{MovingToZeroPosition}, \inlinecordis{MovingToEndPosition}, and \inlinecordis{Disabled}.
Furthermore, it contains subdiagrams \inlinecordis{InZeroPosition}, \inlinecordis{InEndPosition}.
We next elaborate the most important details of these.

\begin{figure}[ht]
\centering
  \centering
  \includegraphics[scale=.4]{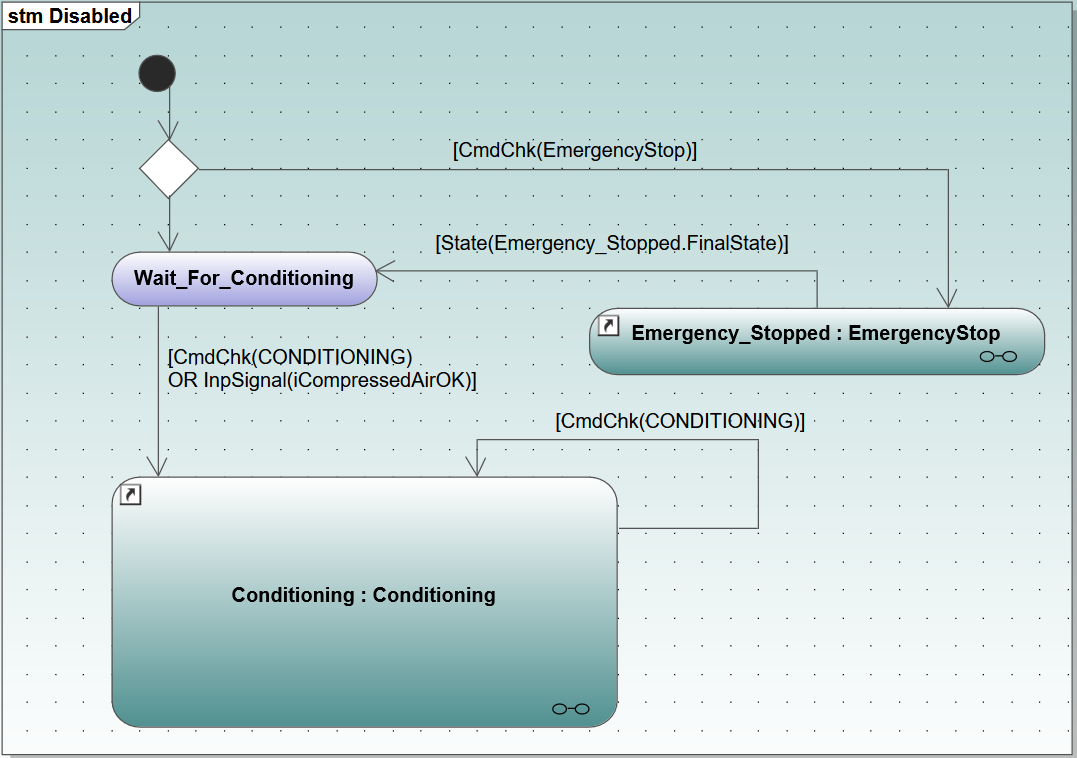}
  \caption{Substatemachine \inlinecordis{Disabled}}
  \label{fig:cylinder-disabled-sm}
\end{figure}

\paragraph{Disabled.}
Initially, the cylinder model enters substatemachine \inlinecordis{Disabled}.
This state machine is shown in Figure~\ref{fig:cylinder-disabled-sm}.

\begin{figure}[ht]
  \centering
  \includegraphics[scale=.5]{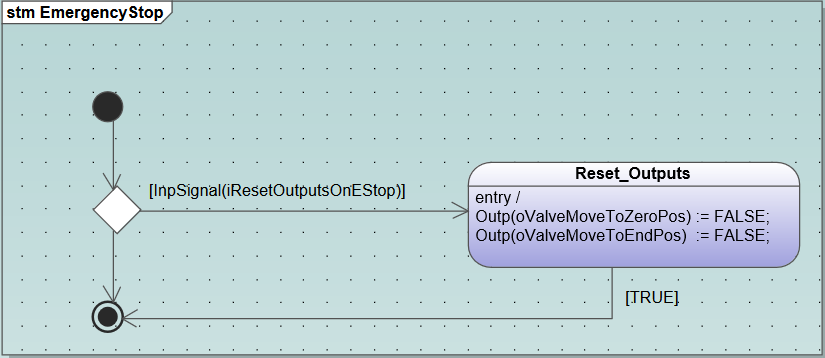}
  \caption{Subdiagram \inlinecordis{EmergencyStop}}
  \label{fig:cylinder-emergency-stop-sm}
\end{figure}

\begin{figure}[ht]
\centering
\includegraphics[scale=.4]{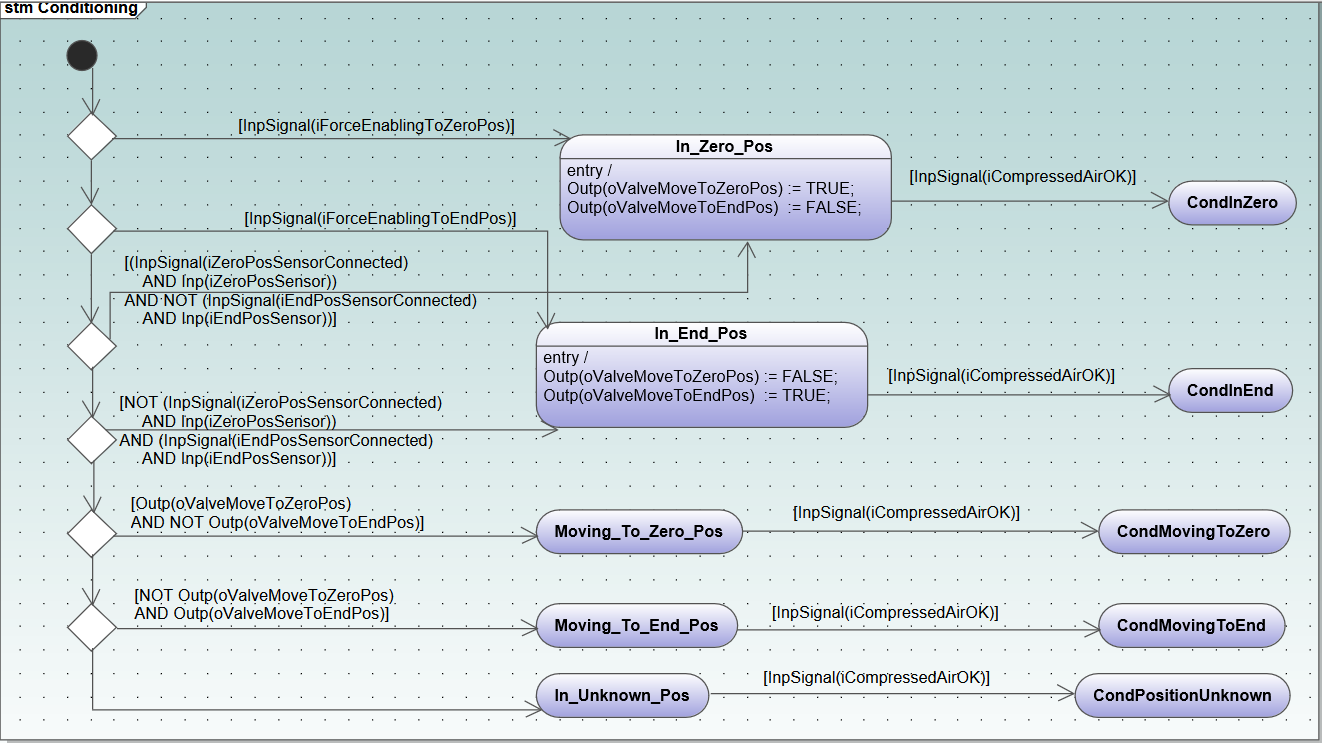}
\caption{State machine diagram \inlinecordis{Conditioning}}
\label{fig:Cond}
\end{figure}

From the initial state of \inlinecordis{Disabled}, if command \inlinecordis{EmergencyStop} was accepted, the system moves to subdiagram \inlinecordis{EmergencyStop}, shown in Figure~\ref{fig:cylinder-emergency-stop-sm}.
In \inlinecordis{EmergencyStop}, if input signal \inlinecordis{iResetOutputsOnEStop} is \true, \inlinecordis{Reset_Outputs} is entered, and outputs \inlinecordis{oValeMoveToZeroPos} and \inlinecordis{oValveMoveToEndPos} are set to \false, stopping all movement of the cylinder.
Once this code has finished executing, the system moves to \inlinecordis{Wait_For_Conditioning}. This state in \inlinecordis{Disabled} is also reached directly from the initial state if no \inlinecordis{EmergencyStop} was issued.
Subdiagram \inlinecordis{Conditioning}, shown in Figure~\ref{fig:Cond}, determines, based on the current values of the inputsignals, inputs and outputs, which of the states in \inlinecordis{Main} reflects the current situation of the cylinder using a cascade of choice nodes. The cascade of choice nodes can be interpreted as an \lstinline[language=c]{if ... else if ... else ...} conditional.
The states without outgoing transitions are used from statemachine \sm{Main} to determine the appropriate exit from \inlinecordis{Disabled}.

\begin{figure}[!h]
\centering
\includegraphics[width=0.88 \linewidth]{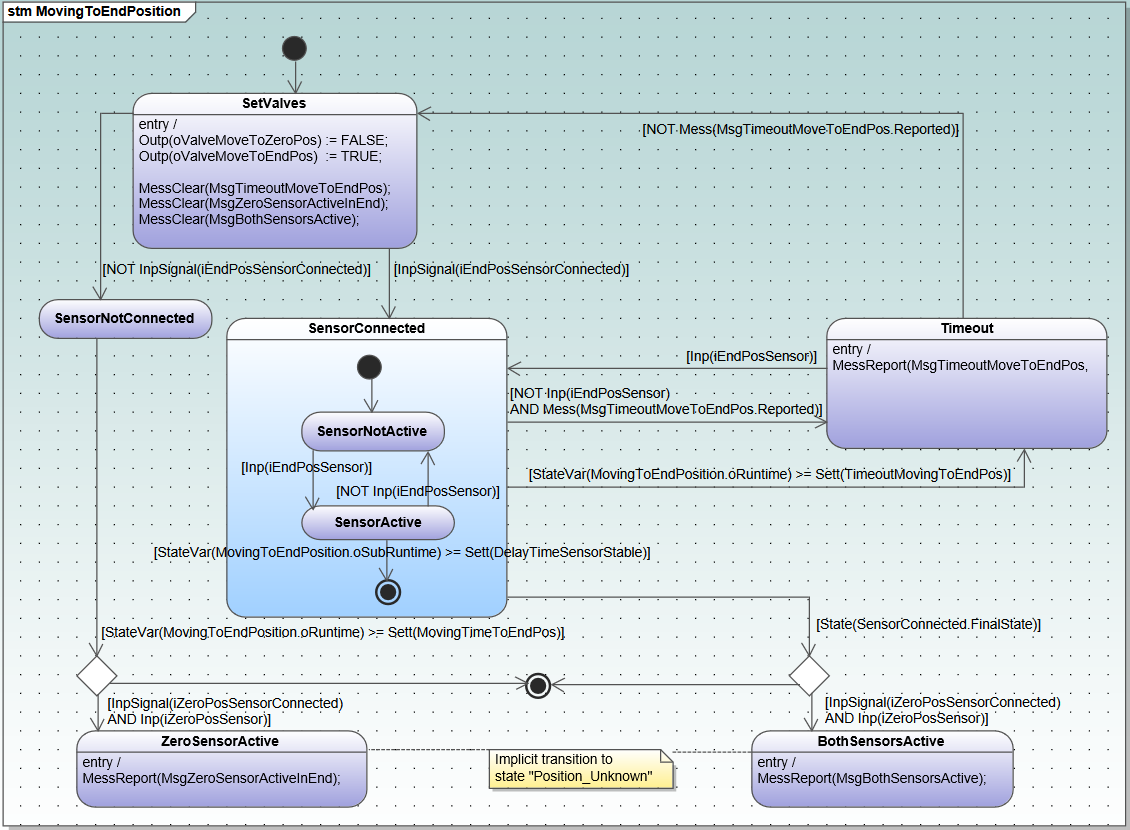}
\caption{Substatemachine \inlinecordis{MovingToEndPosition}}
\label{fig:cylinder-movingtoendposition-sm}
\end{figure}

\paragraph{MovingToEndPosition.}
The two substatemachines \inlinecordis{MovingToEndPosition} and \inlinecordis{MovingToZeroPosition}, are symmetric.
We therefore only explain the first.
The substatemachine is shown in Figure~\ref{fig:cylinder-movingtoendposition-sm}.\footnote{For the sake of completeness, \inlinecordis{MovingToZeroPosition} is included in Appendix~\ref{sec:extra-diagrams}.}

When \inlinecordis{MovingToEndPosition} is entered, first \inlinecordis{SetValves} is reached. Here, the valves are set such that the cylinder starts moving to the end position.
From state \inlinecordis{SetValves}, if input signal \inlinecordis{iEndPosSensorConnected} is \true, composite state \inlinecordis{SensorConnected} is reached.

If \inlinecordis{iEndPosSensor} is \true, a transition can be taken to \inlinecordis{SensorActive}.
This state can be left either if \inlinecordis{iEndPosSensor} becomes \false, or if the sensor value has been stable sufficiently long.
In the latter case, \inlinecordis{SensorConnected.FinalState} is reached.
This triggers the transition to the rightmost choice node in the diagram.
From this node, the transition to \inlinecordis{BothSensorsActive} is taken if not only \inlinecordis{iEndPosSensor} but also \inlinecordis{iZeroPosSensor} is \true\ (and connected); this is an unexpected situation where the hardware indicates the cylinder is in its zero position and its end position at the same time. The \inlinecordis{Main} state machine responds to this error situation through the transitions in Figure~\ref{fig:cylinder-main-sm} with, e.g., guard \inlinecordis{State(Moving_To_End_Pos.BothSensorsActive)}.
If this error situation does not happen, the final state of \inlinecordis{MovingToEndPosition} is reached.
Note that \inlinecordis{SensorConnected} can be left to state \inlinecordis{Timeout} whenever the system takes too long to stably reach state sensor active.

\begin{figure}[ht]
\centering
\includegraphics[width=\linewidth]{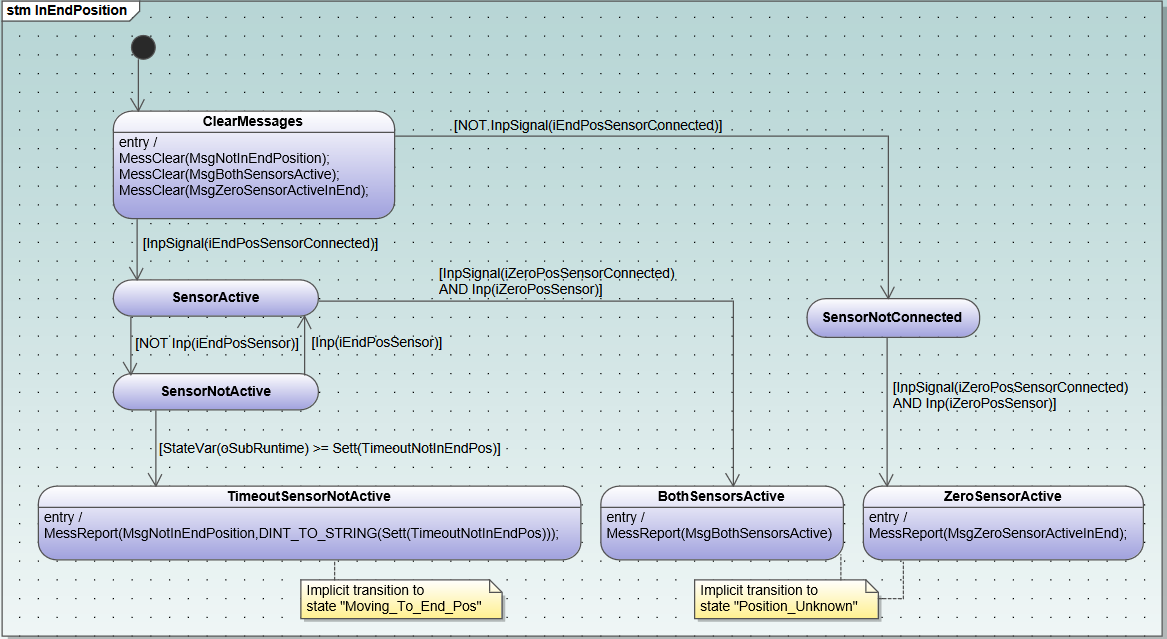}
\caption{State machine diagram \inlinecordis{InEndPosition}}
\label{fig:cylinder-inendposition-sm}
\end{figure}

\paragraph{InEndPosition.}
Subdiagrams \inlinecordis{InEndPosition} and \inlinecordis{InZeroPosition} are also symmetric, and describe the behavior of the cylinder when it is completely extended or completely retracted, respectively.
We here only describe the behavior of \inlinecordis{InEndPosition} shown in Figure~\ref{fig:cylinder-inendposition-sm}.\footnote{For the sake of completeness, \inlinecordis{InZeroPosition} is included in Appendix~\ref{sec:extra-diagrams}.}

Essentially this subdiagram keeps track of the current value of the input \inlinecordis{iEndPosSensor}, moving to state \inlinecordis{SensorActive} if the \inlinecordis{iEndPosSensor} is \true, and to \inlinecordis{SensorNotActive} otherwise.
The remainder of the diagram covers different error situations and their response.
Whenever the diagram is in state \inlinecordis{TimeoutSensorNotActive} the logic in \inlinecordis{Main} in Figure~\ref{fig:cylinder-main-sm} can take a transition to \inlinecordis{Moving_To_End_Pos}.
Similarly, if the diagram reaches \inlinecordis{BothSensorsActive} or \inlinecordis{ZeroSensorActive}, state machine \inlinecordis{Main} ensures that a transition to state \inlinecordis{Position_Unknown} is taken.

\section{Model checking Cordis models using mCRL2}\label{sec:mcrl2translation_verification}

We enable formal verification of Cordis models through an automatic translation of the Cordis semantics to the modelling language mCRL2~\cite{groote2014modeling}.
The language is based on proces algebra with data.
Its associated tool set~\cite{mCRL2} can be used for modelling, validating and verifying systems.
Although mCRL2 allows the specification of communicating, parallel processes, the formalization of the semantics of Cordis model we present in this work only uses sequential processes.
In the following subsections we describe our translation to mCRL2, see Section~\ref{sec:mcrl2-translation}, and the formalization of a number of properties, see Section~\ref{sec:formalization-properties}.
We again use the model of a pneumatic cylinder (see Example~\ref{ex:cylinder-class}) as a running example.

\subsection{Translation to mCRL2}\label{sec:mcrl2-translation}

The mCRL2 specification of Cordis models consists of a sequence of several processes that model the behavior of the system.
Essentially, each execution step depicted in Figure~\ref{fig:execution-order} is represented by a process in the mCRL2 specification.

The basic building block in a process is an action, such as \inlinemcrl{a}, \inlinemcrl{b}, that can be parameterized with data, e.g., \inlinemcrl{a(0)}, \inlinemcrl{b(false)}.
When \inlinemcrl{p} and \inlinemcrl{q} are processes, the sequential composition \inlinemcrl{p.q} denotes the process in which first \inlinemcrl{p} is executed, and upon termination, \inlinemcrl{q} is executed.
The alternative composition, or choice, \inlinemcrl{p + q} denotes that either \inlinemcrl{p} or \inlinemcrl{q} is executed.
Recursive processes can be defined by writing process equations of the form \inlinemcrl{P = q}, where \inlinemcrl{P} is the name of the process, and \inlinemcrl{q} is a process expression in which named processes are referred to in the same way as actions.

In the mCRL2 specification of Cordis models, all processes are parameterized with the current configuration of the system, i.e. the current states of all state machines, and the current values of all class properties and operations (commands).
In what follows we sometimes omit (part of) the parameters, and write \inlinemcrl{...} instead.
Each state machine in the model is identified uniquely by a nonnegative index.

\begin{example}
For the cylinder example discussed in this paper, the process describing the top-level of the system is as follows.
\begin{lstlisting}[language=mcrl2]
P_main(state_machine: Nat, s1: List(State), ..., cmd2: Command, cmd2_ready: Bool,
              cmd2_accepted: Bool, behaviors: List(Int), ...,
              M2'ToggleToZeroPosition: Bool, M2'iZeroPosSensor: Bool,
              M2'iEndPosSensor: Bool, M2'oValveMoveToZeroPos: Bool,
              M2'oValveMoveToEndPos: Bool, M2'iCompressedAirOK: Bool,
              M2'iZeroPosSensorConnected: Bool, M2'iEndPosSensorConnected: Bool,
              M2'iResetOutputsOnEStop: Bool, M2'iForceEnablingToZeroPos: Bool,
              M2'iForceEnablingToEndPos: Bool, M2'oInZeroPosition: Bool,
              M2'oInEndPosition: Bool, M2'oEnabled: Bool, ...) = P_set_inputs();
\end{lstlisting}
\end{example}
The state machine that is currently executing is tracked by the parameter \inlinemcrl{state_machine}, which is a natural number.
The current configuration of the system is tracked by, for every top-level statemachine, keeping a list of states. This contains all states that are contained in the top-level state machine and the subdiagrams and substatemachines it refers to, that are currently active.
States are defined using \inlinemcrl{sort State = struct State_(state: Nat, entry: List(behavior),} \inlinemcrl{cont: List(behavior))} such that each state contains its unique identifier \inlinemcrl{state}, and its entry and continuous behavior, \inlinemcrl{entry} and \inlinemcrl{cont}.
For every machinepart \inlinemcrl{i}, again indexed by an integer, the command that is currently on the interface (along with some additional information) is kept in \inlinemcrl{cmdi}. In the example, the machinepart \inlinemcrl{Cylinder} has index $2$.
Furthermore, the class properties are stored as parameters. For instance, class property \inlinemcrl{iEndPosSensor}, an input of the machinepart with index $2$, is stored as \inlinemcrl{M2'iEndPosSensor}, were \inlinemcrl{M2} refers to machinepart $2$.

We next explain the parts of the mCRL2 specification that are most relevant for the Cylinder model, following the execution of the Cordis semantics.

At the beginning of each cycle, the values of the inputs are received by the system.
As the inputs are not controlled by the system, we model these by receiving arbitrary values of the domain of the inputs.
\begin{example}
For the cylinder model this is formalized as follows.
\begin{lstlisting}[language=mcrl2]
P_set_inputs(..., M2'iZeroPosSensor: Bool, M2'iEndPosSensor: Bool, ...) =
        sum M2'iZeroPosSensor', M2'iEndPosSensor': Bool
        . inputs(M2'iZeroPosSensor', M2'iEndPosSensor')
  . P_set_free_input_signals(M2'iZeroPosSensor = M2'iZeroPosSensor',
                                                       M2'iEndPosSensor = M2'iEndPosSensor');
\end{lstlisting}

\end{example}
In this equation \inlinemcrl{P_set_inputs} is a parameterized process. We focus in particular on the \stereotype{Input} parameters from the class diagram.
The \inlinemcrl{sum} denotes a generalized alternative composition. Using bound variables \inlinemcrl{iEndPosSensor'} and \inlinemcrl{iZeroPosSensor'}, it generates the choice between all four combinations of values to the action \inlinemcrl{inputs}.
Subsequently, \inlinemcrl{P_set_free_input_signals} is called, where the new values of the inputs are assigned to the process parameters.
As the inputs are read in every cycle of the execution, a liveness analysis can be performed, setting the value of the input parameters to a default value if it will not be used in the current state.
The process \inlinemcrl{P_set_free_input_signals} is  similar to \inlinemcrl{P_set_inputs}. For the sake of modelling, this allows to set arbitrary values to the input signals.

This process in turn calls \inlinemcrl{P_set_free_commands}, which cycles through all machineparts to model commands that are issued by the environment.
Like states, commands are indexed by a natural number. Issuing commands is modeled by non-deterministic choice over the commands of the machinepart.  If no new command is issued this is indicated using action \inlinemcrl{no_freecmd}.
\begin{example}
For the cylinder, an outline of the process that sets free commands is the following.
\begin{lstlisting}[language=mcrl2]
P_set_free_commands_2(..., cmd2: Command, ...) =
    freecmd(6) . P_machineparts(cmd2 = M2'MOVE_TO_ZERO_POSITION,
                                                            cmd2_ready = false, cmd2_accepted = false)
  + ...
  + no_freecmd . P_machineparts()
\end{lstlisting}
\end{example}

Once all external inputs to the system have been established, the cyclic execution of machineparts is performed.
In the case of the cylinder, only machinepart 2 needs to execute.
First, the prestate is executed (which is empty in case of the cylinder). Subsequently, the command on the interface is executed.

\begin{example}
For instance, command \inlinemcrl{MOVE_TO_ZERO_POSITION} has index $6$ in the cylinder model. It is executed using the following code.
\begin{lstlisting}[language=mcrl2]
P_command_6(..., s1: List(State), ..., cmd2: Command,
                   cmd2_ready: Bool, cmd2_accepted: Bool, ...) =
    (isCommand2_MOVE_TO_ZERO_POSITION(cmd2) && !cmd2_accepted)
        -> command(6, true)
            . P_command_6_exec(behaviors = accept(cmd2), cmd2_accepted = true,
                                cmd2_ready = S79 in s1 || S103 in s1 || S89 in s1 || S93 in s1)
  + (isCommand2_MOVE_TO_ZERO_POSITION(cmd2) && cmd2_accepted)
        -> chk_ready . P_statemachines_M2(cmd2_ready = ...);
\end{lstlisting}
\end{example}
If the command is new on the interface, \inlinemcrl{cmd2_accepted} is currently \inlinemcrl{false}. If the command guard evaluates to true, the second argument of the command action is \inlinemcrl{true}, indicating the command is accepted, and otherwise it is \inlinemcrl{false}.
If the command is accepted, the command accept behavior is listed for execution, indicated by \inlinemcrl{behaviors = accept(cmd2)}; if the command is rejected, \inlinemcrl{reject(cmd2)} is assigned instead.
Furthermore, the process records whether the command was accepted, and the command ready condition is evaluated in the assignment to \inlinemcrl{cmd2_ready}.
In this case, the command ready condition is \true\ if the state machine is currently in one of four states.
If the command was accepted in a previous cycle, only the command ready condition is checked and its valuation is updated in the same way as in the case of a new command.
If the command was not new, the state machines, in order, get the turn to take one transition each. If a new command was accepted, first the corresponding behaviors are executed in \inlinemcrl{P_command_6_exec}, before moving on to the state machine execution.

To execute the state machines, there is a separate process for each top-level state machine. For instance, in the cylinder model, statemachine \inlinecordis{Main} has index $1$, and the corresponding process is \inlinemcrl{P_statemachine_S1}.
The process cycles through all statemachines that are ancestors of the top-level state machine in order, and allows each to take a transition.
Transitions are defined by \inlinemcrl{sort Transition = struct Transition_(source:List(State),} \inlinemcrl{dest:List(State), behavior:List(behavior))}
that is, its source states \inlinemcrl{source}, its target states \inlinemcrl{dest}, and the behavior that is executed when taking the transition, \inlinemcrl{behavior}.
The statemachine process offers a non-deterministic choice over all transitions in the statemachines.

\begin{example}
We give an example of one transition in the process of the cylinder. The other transitions are similar.
\begin{lstlisting}[language=mcrl2]
P_transitions_S1(state_machine: Nat, s1: List(State),...) =
    ...
  + (state_machine == 1 && (head(source(t100)) in s1)
         && (!M2'iCompressedAirOK || isCommand2_EmergencyStop(cmd2) && cmd2_accepted))
     -> trans(100)
        .P_execute_behaviors_S1(behaviors = behavior(t100) ++ entry(head(dest(t100))),
                                 s1 = dest(t100) ++ remove_prefix(s1, rhead(source(t100))), ...)
  + ...
\end{lstlisting}
In this excerpt, \inlinemcrl{t100} refers to the transition with source state \inlinecordis{Main.Enabled} and target state \inlinecordis{Main.Disabled.InitialState} in Figure~\ref{fig:cylinder-main-sm} which is guarded by \inlinecordis{[NOT InpSignal(iCompressedAirOK) OR CmdChk(EmergencyStop)]}.
\\The summand consists of a guard which says that statemachine  \inlinecordis{Main} is executing, i.e., \inlinemcrl{state_machine == 1}, and source state \inlinecordis{Main.Enabled} is part of the current configuration, i.e., \inlinemcrl{(head(source(t100)) in s1)}. Furthermore, it checks if command \inlinecordis{EmergencyStop} was accepted using \inlinemcrl{isCommand2_Emergency}\inlinemcrl{Stop(cmd2) && cmd2_accepted}.\footnote{Note that in mCRL2, \inlinemcrl{&&} (conjunction) binds stronger than \inlinemcrl{||} (disjunction).}
In case the condition is satisfied, the action \inlinemcrl{trans(100)} is executed and \inlinemcrl{P_execute_behaviors_S1} is called in order to execute the behaviors labelling the transition (if any), \inlinemcrl{behavior(t100)}, as well as the entry behavior of the target state, \inlinemcrl{entry(head(dest(t100)))}.
The next state reached in the statemachine is \inlinemcrl{dest(t100) ++ remove_prefix(s1, rhead(source(t100)))}, where \inlinemcrl{dest(t100)} describes the configuration that is reached after taking the transition, and \inlinemcrl{remove_prefix(s1, rhead(source(t100)))} removes all the states that are left by taking the transition from the configuration.
\end{example}
If multiple transitions are enabled on different levels in the hierarchy, for instance both a state and its strict descendant could take a transition based on the current configuration and the guards, priority is given to the transition from the state highest in the hierarchy. For transitions from substates and substatemachines this is encoded in the mCRL2 translation by including, in the condition generated for the transition, the negation of the guard of all transitions that take priority.

After all state machines have executed one transition and the corresponding behavior, the poststate is executed. To this end, once the last state machine's behavior has executed, process \inlinemcrl{P_poststate_M2} is called, with \inlinemcrl{behaviors = poststate(1)}, i.e., the behaviors that correspond to the poststate of top-level state machine \inlinecordis{Main}.

The execution of behaviors for the prestate, poststate and the transitions are all executed in a similar fashion. We here explain the execution for the poststate where behaviors are coded using structured text. The semantics of structured text is translated to a sequence of mCRL2 processes, whose parameter updates capture the semantics.
The structured text itself is translated to a list of such behaviors.
Executing the behavior then amounts to looping through the list of behaviors, and executing the corresponding processes.

\begin{example}
For the poststate of the cylinder, this is modelled as follows.
\begin{lstlisting}[language=mcrl2]
P_poststate_M2(..., behaviors: List(Int), ...) =
    (behaviors == []) -> post_done.P_remove_command_M2()
  + (behaviors != [] && head(behaviors) == 3)
      -> post(3).P_3(behaviors = tail(behaviors));
\end{lstlisting}
The actual updates done in \inlinemcrl{P_3} are, in this case, simply updates to process parameters, reflecting the assignments in the poststate. For the sake of brevity we omit the details.
\end{example}
Once all behaviors have been executed, transition \inlinemcrl{post_done} is taken, and if a command was on the interface and the command ready condition was true, in process \inlinemcrl{P_remove_command_M2}, the command is removed from the interface, and \inlinemcrl{cmd2_accepted} and \inlinemcrl{cmd2_ready} are reset to their default value.
Execution subsequently repeats from the beginning.

\subsection{Formal verification of requirements}\label{sec:formalization-properties}
One of the primary goals of formalizing Cordis models using mCRL2 is to enable the formal verification of requirements.
In this section, we first describe the requirements. Subsequently we discuss their formalization.

\subsubsection{Requirements}
In total, we have formulated 12 requirements for the cylinder, and formalized and verified them. In this section, we describe two of these requirements, one safety requirement and one liveness requirement. The remaining requirements and their formalizations are included in Appendix~\ref{sec:extra-requirements}.

The requirements we consider are the following two:
\begin{enumerate}
    \item \label{req:if-oInEndPosition-or-oInZeroPosition-then-oEnabled} Invariantly, if one of the output signals \inlinecordis{oInEndPosition} or \inlinecordis{oInZeroPosi-} \inlinecordis{tion} is \true, also output signal \inlinecordis{oEnabled} is \true.

    \item \label{req:if-not-oEnabled-and-iCompressedAirOk-then-oEnabled} Whenever output signal \inlinecordis{oEnabled} is \false\ and input signal \inlinecordis{iCompressedAir} \inlinecordis{OK} is \true, inevitably output signal \inlinecordis{oEnabled} becomes \true\ unless command \inlinecordis{CONDITIONING} is accepted.
\end{enumerate}

\subsubsection{Formalization of requirements using the modal $\mu$-calculus}\label{sec:mu-calculus-formulas}
We describe requirements using the first order modal $\mu$-calculus~\cite{GM1999}.
This is a very expressive temporal logic that extends the $\mu$-calculus with data.

In general, the requirements we are interested in refer to the interfaces of the machine parts, that is, their inputs, input signals, commands, output signals, and outputs. The first three are visible in the transitions of the process we generate: they are set explicitly. However, the output signals and outputs are only available implicitly as part of the process parameters.
In order to expose their values, we have extended the translation with self-loops that are labelled with transitions that show the current value of the outputs and output signals. For this, we use actions such as \inlinemcrl{state_M2'oInEndPosition(true)}, where \inlinemcrl{state} indicates this is a state loop, \inlinemcrl{M2} refers to the machinepart, \inlinemcrl{oInEndPosition} is the name of the output, and \inlinemcrl{true} is the current value of the output. Using self-loops labelled with action \inlinemcrl{states}, we also expose the current state of the system.

This is used to formalize the first requirement as follows.
\begin{lstlisting}[language=mcrl2]
[true*](<state_M2'oInEndPosition(true)||state_M2'oInZeroPosition(true)>true
                    => <state_M2'oEnabled(true)>true)
\end{lstlisting}
This formula should be read as follows.
First, \inlinemcrl{[true*]} represents all sequences consisting of zero or more actions. After each such sequence, i.e., in every reachable state of the system, the remainder of the formula should hold.
For the remainder, note that formula \inlinemcrl{<a>true} holds in every state with an outgoing \inlinemcrl{a} transition. If we write action formula \inlinemcrl{a || b} inside a modality, this matches the set of actions containing \inlinemcrl{a}, \inlinemcrl{b}; essentially, \inlinemcrl{||} here denotes the union of the sets of action represented by \inlinemcrl{a} and \inlinemcrl{b}, which are the singleton sets containing \inlinemcrl{a} and \inlinemcrl{b}, respectively.
Hence, \inlinemcrl{<state_M2'oInEndPosition(true)||state_M2'oInZeroPosition(true)>true} holds in every state that has an outgoing transition labelled \inlinemcrl{state_M2'oInEndPosition(true)}, or it has a transition labelled \inlinemcrl{state_M2'oInZeroPosition(true)}. In each such state, the formula requires that also \inlinemcrl{<state_M2'oEnabled(true)>true} holds, i.e., the state has an outgoing transition labelled \inlinemcrl{state_M2'oEnabled(true)}.
We refer to~\cite{groote2014modeling} for a more extensive introduction to the $\mu$-calculus.

The second requirement is formalized as follows.
\begin{lstlisting}[language=mcrl2]
nu X(enabled: Bool = false, compressedAirOk: Bool = false) .
    (forall e: Bool . <state_M2'oEnabled(e)>true =>
        ((forall c: Bool . [exists a2, a3, a4, a5, a6: Bool .
            free_input_signals(c, a2, a3, a4, a5, a6)]X(e,c)) &&
            [!exists a1, a2, a3, a4, a5, a6: Bool .
                free_input_signals(a1, a2, a3, a4, a5, a6)]X(e,compressedAirOk))) &&
    (forall e: Bool . [state_M2'oEnabled(e)]false =>
        ((forall c: Bool . [exists a2, a3, a4, a5, a6: Bool .
            free_input_signals(c, a2, a3, a4, a5, a6)]X(enabled,c)) &&
            [!exists a1, a2, a3, a4, a5, a6: Bool .
                free_input_signals(a1, a2, a3, a4, a5, a6)]X(enabled,compressedAirOk))) &&
    (val(!enabled && compressedAirOk) =>
        mu X . [!((exists a2, a3, a4, a5, a6: Bool .
                                    free_input_signals(false, a2, a3, a4, a5, a6)) ||
                                    command(9, true) ||
                            (exists b: Bool . state_M2'oValveMoveToZeroPos(b) ||
                                                                state_M2'oValveMoveToEndPos(b) ||
                                                                state_M2'oInZeroPosition(b) ||
                                                                state_M2'oInEndPosition(b) ||
                                                                state_M2'oEnabled(b)) ||
                            (exists i: Nat, l: List(Nat) . states(i, l))
        )]X || <state_M2'oEnabled(true)>true)
\end{lstlisting}
This formula uses a greatest fixed point (\inlinemcrl{nu}) and a least fixed point (\inlinemcrl{mu}). The greatest fixed point is parameterized by two Boolean variables, \inlinemcrl{enabled} and \inlinemcrl{compressedAirOk}, that keep track of whether \inlinemcrl{oEnabled} or \inlinemcrl{iCompressedAirOk} \\have become \true, respectively.
In order to keep track of these values, we distinguish two cases. If a transition
\inlinemcrl{state_M2'oEnabled(e)} is enabled, denoted by \inlinemcrl{forall e: Bool. <state_M2'oEnabled(e)>true}, we check if an action \inlinemcrl{free_input_signals} is enabled. If so, we determine the value assigned to \inlinecordis{iCompressedAirOk} using \inlinemcrl{forall c: Bool . [exists a2, a3, a4, a5, a6: Bool .free_input_signals(c, a2, a3, a4, a5, a6)]X(e,c))}. We use \inlinemcrl{exists} inside the modality to represent generalised union, and match any value for the rest of the input signals. We update  \inlinemcrl{enabled} to the value observed by the self-loop, and \inlinemcrl{compressedAirOk} to the value set in \inlinemcrl{free_input_signals}.
If \inlinemcrl{free_input_signals} is not enabled, only \inlinecordis{compressedAirOk} is updated.
The case where \inlinemcrl{state_M2'oEnabled(e)} is not enabled is handled in a similar way.
That is, if an action \inlinemcrl{free_input_signals} is enabled, we only update the value of \inlinemcrl{compressedAirOk} by observing the values in \inlinemcrl{free_input_signals}. Otherwise, the value of the parameters of \inlinemcrl{X} are unchanged.
To summarize, the first part of the formula describes an invariant property, and it keeps track of the most recent value of \inlinemcrl{oEnabled} and \inlinemcrl{iCompressedAirOK}.

Now, if \inlinemcrl{enabled} is \false, and \inlinemcrl{compressedAirOk} is \true,
the least fixed point subformula needs to hold.
To interpret this formula, we first note that formula \inlinemcrl{mu Y . [!a]Y || <b>true} captures that inevitably a state is reached where a \inlinemcrl{b} transition is enabled, unless an \inlinemcrl{a} transition happens.
So, in principle, the following formula denotes that, as long as \inlinemcrl{iCompressedAirOk}, does not become \inlinemcrl{false}, represented by the first argument to \inlinemcrl{free_input_signals}, and command \inlinemcrl{CONDITIONING} is not accepted, represented by \inlinemcrl{command(9, true)}, then we inevitably end up in a state where \inlinemcrl{oEnabled} is \true.
\begin{lstlisting}[language=mcrl2]
    mu Y . [!((exists a2, a3, a4, a5, a6: Bool .
                            free_input_signals(false, a2, a3, a4, a5, a6)) ||
                        command(9, true))]Y ||
                 <state_M2'oEnabled(true)>true)
\end{lstlisting}
However, as we extended the model with self-loops to expose the current value of output signals, by taking such self-loops we trivially end up in an infinite sequence on which no state where \inlinemcrl{oEnabled} holds is reached. We therefore need to exclude paths through these self-loops.\footnote{We here rely on the fact that the additional information is only exposed through self-loop transitions. This avoids the need for introducing an additional greatest fixed point.}

\subsection{Results}\label{results}
We have verified the two properties from Section~\ref{sec:formalization-properties}, as well as 10 additional requirements.
For our experiments we have used Cordis Modeler version 3.14.1630. 7156 and mCRL2 tool set Release 202106.0.
The cylinder model described and studied in this paper is relatively simple. This is reflected by the verification time: each of the properties can be verified in less than 5 seconds.
Property~\ref{req:if-not-oEnabled-and-iCompressedAirOk-then-oEnabled} is \false, and all of the other requirements are satisfied by the model.
In case a property does not hold, the mCRL2 tool set offers a counterexample. In the next section, we discuss the counterexample to property~\ref{req:if-not-oEnabled-and-iCompressedAirOk-then-oEnabled} in detail.

\section{Discussion}\label{sec:discussion}
As observed in the previous section, requirement~\ref{req:if-not-oEnabled-and-iCompressedAirOk-then-oEnabled} does not hold for the Cylinder model.
The mCRL2 tool set provides a counterexample that is a subset of the labelled transition system that underlies the specification of the cylinder, and that contains sufficient information to prove that the property is violated~\cite{DBLP:conf/cade/WesselinkW18}.

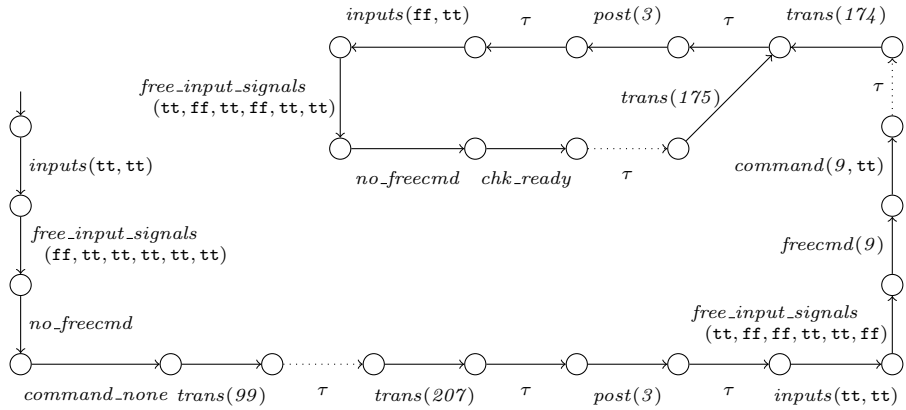
\begin{figure}
    \centering
\begin{tikzpicture}[node distance=30pt]
\scriptsize
  \node [state, initial above] (1) {};
  \node [state, below of=1] (2) {};
  \node [state, below of=2] (3) {};
  \node [state, below of=3] (4) {};
  \node [state, right =1.7cm of 4] (5) {};
  \node [state, right =of 5] (6) {};
  \node [state, right =of 6] (7) {};
  \node [state, right =of 7] (12) {};
  \node [state, right =of 12] (13) {};
  \node [state, right =of 13] (14) {};
  \node [state, right =of 14] (15) {};
  \node [state, right =1.2cm of 15] (16) {};
  \node [state, above of=16] (17) {};
  \node [state, above of=17] (18) {};
  \node [state, above of= 18] (19) {};
  \node [state, above of= 19] (20) {};
  \node [state, left =1.2cm of 20] (26) {};
  \node [state, left =of 26] (27) {};
  \node [state, left =of 27] (28) {};
  \node [state, left =of 28] (29) {};
  \node [state, left =1.5cm of 29] (30) {};
  \node [state, below =of 30] (31) {};
  \node [state, right =1.5cm of 31] (32) {};
  \node [state, right =of 32] (33) {};
  \node [state, right =of 33] (34) {};

  \draw[->] (1) edge  node[midway,right] {$\mathit{inputs}(\ttrue,\ttrue)$} (2)
            (2) edge  node[midway,right=-4mm]  {$\begin{array}{c} \mathit{free\_input\_signals} \\ \quad\quad (\ffalse,\ttrue,\ttrue,\ttrue,\ttrue,\ttrue) \end{array}$} (3) 
            (3) edge  node[midway,right] {$\mathit{no\_freecmd}$} (4)
            (4) edge  node[midway,below=2mm] {$\mathit{command\_none}$} (5)
            (5) edge  node[midway,below=2mm] {$\mathit{trans(99)}$} (6)
            (6) edge[dotted]  node[midway,below=2mm] {$\tau$} (7)
            (7) edge  node[midway,below=2mm] {$\mathit{trans(207)}$} (12)
            (12) edge  node[midway,below=2mm] {$\tau$} (13)
            (13) edge  node[midway,below=2mm] {$\mathit{post(3)}$} (14)
            (14) edge  node[midway,below=2mm] {$\tau$} (15)
            (15) edge  node[midway,below=2mm] {$\mathit{inputs}(\ttrue,\ttrue)$} (16)
            (16) edge  node[midway,left=-1mm] {$\begin{array}{c} \mathit{free\_input\_signals} \\ \quad\quad (\ttrue,\ffalse,\ffalse,\ttrue,\ttrue,\ffalse) \end{array}$} (17) 
            (17) edge  node[midway,left] {$\mathit{freecmd(9)}$} (18)
            (18) edge  node[midway,left] {$\mathit{command(9,\ttrue)}$} (19)
            (19) edge[dotted]  node[midway,left] {$\tau$} (20)
            (20) edge  node[midway,above=2mm] {$\mathit{trans(174)}$} (26)
            (26) edge  node[midway,above=2mm] {$\tau$} (27)
            (27) edge  node[midway,above=2mm] {$\mathit{post(3)}$} (28)
            (28) edge  node[midway,above=2mm] {$\tau$} (29)
            (29) edge  node[midway,above=2mm] {$\mathit{inputs}(\ffalse,\ttrue)$} (30)
            (30) edge  node[midway,left=-1mm] {$\begin{array}{c} \mathit{free\_input\_signals} \\ \quad\quad (\ttrue,\ffalse,\ttrue,\ffalse,\ttrue,\ttrue) \end{array}$} (31) 
            (31) edge  node[below=2mm] {$\mathit{no\_freecmd}$} (32)
            (32) edge  node[below=2mm] {$\mathit{chk\_ready}$} (33)
            (33) edge[dotted]  node[midway,below=2mm] {$\tau$} (34)
            (34) edge  node[midway,left] {$\mathit{trans(175)}$} (26)
  ;
\end{tikzpicture}
\captionof{figure}{Counterexample found verifying Property~\ref{req:if-not-oEnabled-and-iCompressedAirOk-then-oEnabled}}
\label{fig:CE}
\end{figure}

The counterexample is a lasso, shown in Figure~\ref{fig:CE}. It consists of a sequence of transitions that leads to a cycle on which \sm{iCompressedAirOK} remains true, and command \sm{CONDITIONING} is never accepted, but, \sm{oEnabled} remains \false.
In subsection~\ref{sec:state-machines}, we have described the state machines execution, this same structure is found in the counterexample. In each cycle, after having updated inputs and commands, the state machines are executed in turns in the predetermined order: \sm{Main}, \sm{MovingToZeroPosition}, \sm{MovingToEndPosition}\ and \sm{Disabled}.

For the sake of readability, in Figure~\ref{fig:CE}, the actions which are not essential to describe the trace are labeled with $\tau$; a sequence of $\tau$ labeled transitions is denoted by a dotted arrow labeled $\tau$. We denote \true\ and \false\ as $\mathtt{tt}$ and $\mathtt{ff}$ respectively.

The execution of the counterexample starts from the state with an unlabeled arrow pointing to it.
At the beginning of the first cycle, action $\mathit{inputs(\ttrue,\ttrue)}$ indicates that inputs \sm{iZeroPosSensor} and \sm{iEndPosSensor} are both set to \true; input signal \sm{iCompressedAirOk} is set \false\ while the other input signals are set to \true, and no command is issued.
The first cycle continues with the execution of action $\mathit{trans(99)}$. This indicates that in state machine \sm{Main} the transition from \sm{Main.InitialState} to \sm{Main.Disabled.InitialState} is taken.
As \sm{MovingToZeroPosition} and \sm{MovingToEndPosition} are not currently active, they do not take any transition, indicated by the dotted $\mathit{\tau}$ transition.
Subsequently, state machine \sm{Disabled} takes the transition from \sm{Main.Disabled.InitialState} to \sm{Main.Disabled.Wait\_For\_Conditioning}, denoted by action $\mathit{trans(207)}$.
The poststate concludes the first cycle.

In the second cycle, the inputs remains \true, \sm{iCompressedAirOK} is set to \true\ and command \sm{CONDITIONING} is issued and accepted, expressed by actions $\mathit{freecmd(9)}$ and $\mathit{command(9, \ttrue)}$, respectively. It follows that state machines \sm{Main}, \sm{MovingToZeroPosition}, and \sm{MovingToEndPosition} do not perform transitions.
In state machine \sm{Disabled} action $\mathit{trans(174)}$ is executed, that is, the transition from state \sm{Main.Disabled.Wait\_For\_Conditioning} to state \sm{Main.Disabled.Conditioning.InitialState} is taken.
The remainder of this cycle executes the poststate, shown in the loop moving in counterclockwise direction.

In the third cycle (and subsequent cycles) input signal \sm{iCompressedAirOK} remains \true. No new command is issued, but the action $\mathit{chk\_ready}$ expresses that command \sm{CONDITIONING}, issued in the second cycle, is still active on the interface, and is not yet ready.
State machines \sm{Main}, \sm{MovingToZeroPosition}, and \sm{MovingToEndPosition} do not perform transitions.
Finally, in state machine \sm{Disabled} action $\mathit{trans(175)}$ is executed. That is, the transition from state \sm{Main.Disabled.Conditioning} to \sm{Main.Disabled.Conditioning.InitialState} is taken. In the cylinder model this is the self-transition from and to subdiagram \sm{Conditioning} in Figure~\ref{fig:cylinder-disabled-sm}. The subsequent cycles behave identically, such that state \sm{Main.Disabled.Conditioning.InitialState} is infinitely often re-entered.

\paragraph{Reproducing the counterexample.}
To ascertain that the problem identified by our verification also exists in the implementation, we have analyzed a running implementation of the cylinder model.
Recall that from a Cordis model PLC code can be generated and executed.
By loading the executing PLC code in a debugger, the implementation can be simulated by stepping through the code.
In this way, setting the appropriate inputs and input signals, and issuing the command \inlinemcrl{CONDITIONING} in accordance to the counterexample, we have been able to reproduce the exact behavior of the counterexample.

This is interesting in two respects. First of all, this increases the confidence in the correctness of the translation of the Cordis model semantics to mCRL2. Second, the counterexample contains sufficient information to efficiently reproduce the counterexample in the running system.

\paragraph{Root cause analysis.}
Once it has been established that the counterexample can be reproduced, we analyze it in the context of the cylinder model to understand which part of the model causes this issue.

The system is able to loop in the self-transition from and to subdiagram \sm{Conditioning} in Figure~\ref{fig:cylinder-disabled-sm} because of two reasons: (1) in Cordis models, the outermost enabled transition gets priority over transitions that are more deeply nested in the model, and (2) the command continuously remains active on the interface.
Hence, even if another transition, more deeply nested in subdiagram \sm{Conditioning}, would have been enabled, the self-transition always takes priority.

As the first reason is an inherent part of the semantics of Cordis models, we focus on the transition being enabled in the remainder of the analysis.
Command \sm{CONDITIONING} has
guard condition \sm{State(Main.Disabled)}, command ready condition \sm{NOT State(Main.Disabled)}, and no accept and reject actions.
Thus, command \sm{CONDITIONING} is accepted if the system currently is in state machine \sm{Disabled}, and the command is ready if the system leaves state machine \sm{Disabled}.

In the counterexample, when command \sm{CONDITIONING} is accepted, we are indeed in state machine \sm{Disabled}, so this condition is satisfied, and subdiagram \sm{Conditioning} is entered. As \sm{Conditioning} is a subdiagram of \sm{Disabled}, so the command ready condition is not satisfied in this state, and due to the priorities none of the transitions in Figure~\ref{fig:Cond} can be taken.

\paragraph{Solution.}
Based on the analysis, we observe that issuing the \sm{CONDITIONING} command does not simply trigger the model once to look at the current state of the system, but instead it behaves like a trigger that always remains high.
The solution to avoid this behavior is to modify the command ready condition from \sm{NOT State(Main.Disabled)} to \sm{true}.
This way, the command will indeed only act as a trigger to enter state machine conditioning: when issued and accepted, command \sm{CONDITIONING} stays active on the interface for exactly one cycle.

This change does not affect the relevant behavior of the cylinder model: if the system is currently in a substate of subdiagram \sm{Conditioning}, it can still re-enter subdiagram \sm{Conditioning} if command \sm{CONDITIONING} is issued and accepted again.

Changing the cylinder model accordingly, and re-verifying the requirements shows that, indeed, requirement~\ref{req:if-not-oEnabled-and-iCompressedAirOk-then-oEnabled} holds.

\section{Concluding remarks}\label{sec:conclusion}
In this paper we have discussed the semantics of Cordis models, an extension of standard UML used for modelling complex machine-control applications, in order to enable the verification of these models.
Even though Cordis models are not primarily designed for the application of formal verification, we were able to characterize and implement an automatic translation to mCRL2. As a proof of concept we have verified the behavior of an industrial cylinder model, formalized a number of requirements for this model, and verified these.
We have shown in this paper that the verification process is effective to find bugs, and that the bugs can be reproduced in the actual system using a debugger.

There are some aspects to the formalization and verification process that we do not report on explicitly in this paper. In particular, using earlier versions of the Cordis modeler and the translation to mCRL2 we have uncovered corner cases that were treated differently in the PLC code generated by the Cordis SUITE and in the mCRL2 model. This has resulted in modifications to both the semantics in the Cordis modeler, improving the stability and quality of the generated PLC code, as well as to the translation to mCRL2 to ensure that the semantics of the mCRL2 models are consistent with the PLC code.
In order to understand and debug such issues, both having clear counterexamples in mCRL2, and the ability to step through the PLC code using a debugger have proven indispensable.

\paragraph{Future work.} Cordis models of complete industrial systems usually consists of multiple interacting objects.
To this end, the translator from Cordis models to mCRL2 has been extended to deal with systems that consist of multiple components.

We are currently expanding our work to deal with more complex industrial models. In particular, we are investigating the use of symbolic model checking techniques~\cite{BCM+1992} to deal with the large state spaces they induce.
Furthermore, compositional model checking~\cite{proenca_compositional_2017} could help in the verification of such large models by incrementally generating state spaces of subsystems, reducing them, and combining them into larger subsystems, prior to verification.
We are additionally investigating improvements to static analysis tools that can optimize mCRL2 models, resulting in smaller state spaces~\cite{groote_computer_2002}, and static analysis techniques for Cordis models such as dead variable analysis.
Finally, in our ongoing collaboration with Cordis, we are integrating model checking into the Cordis SUITE in such a way that mCRL2 is used as a verification back-end, and verification is directly accessible from the Cordis modeller.

\subsubsection*{Acknowledgements}
This work was supported partially by the MACHINAIDE project (ITEA3, No. 18030) and through EU regional development funding in the context of the OP-Zuid program (No. 02541). We thank Wieger Wesselink and Yousra Hafidi for contributions to the development of the mCRL2 translation, and Cordis Automation B.V. for their feedback on earlier versions of this paper.

\bibliographystyle{splncs04}
\bibliography{main}

\appendix
\section{Additional requirements}\label{sec:extra-requirements}

Besides the two requirements described in the paper, we have formalized and verified the 10 requirements presented in this appendix.
All of these requirements hold for the cylinder model described in this paper.

\begin{enumerate}[itemsep=3pt] \addtocounter{enumi}{2}
\item \label{req:if-not-conditioning-and-not-iCompressedAirOk-then-not-enabled}
As long as command \inlinecordis{CONDITIONING} has not been accepted and input signal \inlinecordis{iCompressedAirOK} is \false, output signal \inlinecordis{oEnabled} is \false. \label{itm:1}

This is formalized in the $\mu$-calculus as follows.
\begin{lstlisting}[language=mcrl2]
[!(command(9,true)||(exists iZSC,iESC,iR,iFZ,iFE:Bool.
    free_input_signals(true,iZSC,iESC,iR,iFZ,iFE)))*}]
        [state_M2'oEnabled(true)]false
\end{lstlisting}

The formula should be interpreted as follows.
In the translation, \inlinemcrl{command(9,true)} represents the acceptance of command \inlinecordis{CONDITIONING}.
Input signal \inlinecordis{iCompressedAirOK} is the first parameter of action \inlinemcrl{free_input_signals}, so \inlinemcrl{free_input_signals(true,iZSC,iESC,iR,iFZ,iFE)} expresses that \inlinecordis{iCompressedAirOK}  becomes \true, and the other input signals get an arbitrary value.
The first part of the formula, \inlinemcrl{[!(...)*]}, denotes all executions on which command \inlinecordis{CONDITIONING} is not accepted and on which \inlinecordis{iCompressedAirOK}  does not become \true.
Along each such execution, the value of \inlinecordis{oEnabled} never becomes \true, which is formalized by \inlinemcrl{[state_M2'oEnabled(true)]false}.

\item Whenever command \inlinecordis{EmergencyStop} is accepted, inevitably output signal \inlinecordis{oEnabled} becomes \false\ unless a new command is issued. \label{req:if-emergency-stop-eventually-not-enabled}
In the translation, the acceptance of command \inlinecordis{EmergencyStop} is encoded as \inlinemcrl{command(10,true)}.
We would therefore like to express the requirements as
\begin{lstlisting}[language=mcrl2]
[true*.command(10,true)]mu X.([true]X || <state_M2'oEnabled(false)>true)
\end{lstlisting}
That is, whenever command \inlinecordis{EmergencyStop} is accepted then, output signal \inlinecordis{oEnabled} will eventually become \false.
However, if another command is issued, it overwrites the  \inlinecordis{EmergencyStop} command and removes it from the interface.
As a consequence, \inlinecordis{oEnabled} does not necessarily become \false.
We therefore exclude the situation where the command is overwritten by another command.
Furthermore, the self-loops need to be excluded in a similar way to the formula for requirement~\ref{req:if-not-oEnabled-and-iCompressedAirOk-then-oEnabled} in Section~\ref{sec:formalization-properties}.

Therefore, ultimately the requirement is formalized as follows.
\begin{lstlisting}[language=mcrl2]
[true* . command(10,true)]
    mu X.([!((exists i: Nat, b:Bool. command(i,b)) ||
             (exists b: Bool . state_M2'oValveMoveToZeroPos(b) ||
                         state_M2'oValveMoveToEndPos(b) ||
                         state_M2'oInZeroPosition(b) ||
                         state_M2'oInEndPosition(b) ||
                         state_M2'oEnabled(b)) ||
                     (exists i: Nat, l: List(Nat) . states(i,l)))]X ||
	        <state_M2'oEnabled(false)>true)
\end{lstlisting}

\item \label{req:if-MOVE_TO_END_POSITION-inevitably-oInEndPosition-or-oValveMoveToEndPos}
Whenever the system is enabled and command \inlinecordis{MOVE_TO_END_POSITION} is accepted, inevitably either output signal \inlinecordis{oInEndPosition} becomes \true\ or output \inlinecordis{oValveMoveToEndPos} becomes \true, or the system becomes disabled, unless another command is accepted or rejected.

On a high level, we would like to specify the requirement that says whenever command \inlinecordis{MOVE_TO_END_POSITION} is accepted inevitably either output signal \inlinecordis{oInEndPosition} becomes \true\ or output \inlinecordis{oValveMoveToEndPos} becomes \true, formalized as follows.
\begin{lstlisting}[language=mcrl2]
[true* . command(7,true)]
  mu X. ([true]X ||
	<state_M2'oInEndPosition(true) || state_M2'oValveMoveToEndPos(true)>true))
\end{lstlisting}

In the translation, \inlinemcrl{command(7,true)} encodes the acceptance of command \inlinecordis{MOVE_TO_END_POSITION}.
Hence, the part of the requirement stating ``whenever command \inlinecordis{MOVE_TO_END_POSITION} is accepted'' is encoded as \inlinemcrl{[true*} \inlinemcrl{. command(7,true)]}. The least fixed point subformula expresses that inevitably a state should be reached where either output \inlinecordis{oValveMoveToEndPos} or output signal \inlinemcrl{oInEndPosition} inevitably becomes \true.

A complication that we have to deal with is that the requirement does not hold if either currently the system is not enabled, i.e., \inlinecordis{oEnabled} is \false, or it becomes \false\ before reaching the desired state. In other words, the property is not required to hold if we infinitely often traverse a state where \inlinecordis{oEnabled} is \false.

To this end, we replace the \inlinemcrl{[true* . command(7,true)]} with an outer greatest fixed point parameterized by a Boolean variable that keeps track of whether the system is enabled when the command \inlinecordis{MOVE_TO_END_POSITION} is issued. The inner fixed point in entered whenever the system is enabled, \inlinecordis{oEnabled} is \true, and the command \inlinecordis{MOVE_TO_END_POSITION} is accepted.

The resulting formula is the following.
\begin{lstlisting}[language=mcrl2]
nu Y(enabled: Bool = false) .
  (forall b: Bool . <state_M2'oEnabled(b)>true
                                                                => [!command(7,true)]Y(b)) &&
  ([exists b: Bool . state_M2'oEnabled(b)]false
                                                                => [!command(7,true)]Y(enabled)) &&
  (val(enabled) && <state_M2'oEnabled(false)>true
                                                                => [command(7,true)]Y(false)) &&
  (val(enabled) && [state_M2'oEnabled(false)]false
                                                                => [command(7,true)]
    mu X.
      [!(
    			(exists i: Nat, b: Bool . command(i, b)) ||
				(exists b: Bool . state_M2'oValveMoveToZeroPos(b) ||
										            state_M2'oValveMoveToEndPos(b) ||
										            state_M2'oInZeroPosition(b) ||
										            state_M2'oInEndPosition(b) ||
  										            state_M2'oEnabled(b)) ||
  				(exists i: Nat, l: List(Nat) . states(i, l))
      )]X ||
      <state_M2'oInEndPosition(true) || state_M2'oValveMoveToEndPos(true)>true ||
        <state_M2'oEnabled(false)>true
    )
\end{lstlisting}
In more detail, the formula is structured as follows. The outer greatest fixed point is parameterised by the Boolean variable \inlinemcrl{enabled} to keep track of the value assigned to \inlinemcrl{oEnabled}. The greatest fixed point is left only if \inlinemcrl{enabled} is \true\ and does not immediately become \false, and if command \inlinecordis{MOVE_TO_END_POSITION} is accepted. The first conjunct, \inlinemcrl{forall b:Bool .} \inlinemcrl{<state_M2'oEnabled(b)>true => [!command(7,true)]Y(b))} expresses that, if an action \inlinemcrl{state_M2'oEnabled(b)} can be performed, and a transition other than \inlinecordis{MOVE_TO_END_POSITION} is taken, we update the value of \inlinemcrl{enabled} to the value observed in the self-loop. Similarly, if \inlinemcrl{state_M2'oEnabled(b)} is not enabled, any transition that does not accept command  \inlinecordis{MOVE_TO_END_POSITION} keeps the current value of \inlinemcrl{enabled}. The latter is expressed by the conjunct \inlinemcrl{([exists b: Bool .state_M2'oEnabled(b)]false => [!command(7,true)]Y(enabled))}. The third conjunct expresses that, if \inlinemcrl{enabled} is currently \true\ but the state has a self-loop in indicating \inlinemcrl{oEnabled} has become \false, determined by \inlinemcrl{<state_M2'oEnabled(false)>true}, and if then command \inlinecordis{MOVE_TO_END_POSITION} is accepted, we update the value of \inlinemcrl{enabled} to \false\ and stay in the outer fixed point, as the condition for moving to the end position is not met. Finally, the last conjunct, \inlinemcrl{(val(enabled) && [state_M2'oEnabled(false)]false => [command(7,true)]mu X...}, expresses that, if \inlinemcrl{enabled} is \true\ and \inlinemcrl{oEnabled} has not become \false\ and command \inlinecordis{MOVE_TO_END_POSITION} is accepted, the least fixed point subformula needs to hold.

The least fixed point subformula is an extension of the following formula, where additionally self-loops are excluded. In principle, the following subformula denotes that, if no commands are accepted or rejected, indicated by \inlinemcrl{command(i, b)}, we inevitably end up in a state where \inlinemcrl{oInEndPosition} is \true\ or \inlinemcrl{oValveMoveToEndPos} is \true, or the system becomes disabled, \inlinecordis{oEnabled} is \false.
\begin{lstlisting}[language=mcrl2]
   mu X.
    [!((exists i: Nat, b:Bool . command(i, b)) )]X ||
      <state_M2'oInEndPosition(true) || state_M2'oValveMoveToEndPos(true)>true ||
        <state_M2'oEnabled(false)>true)
\end{lstlisting}
We exclude the acceptance and rejection of other commands since this would affect the behavior of the system in a such a way that, if the fourth conjunct holds, neither \inlinemcrl{oInEndPosition} or \inlinecordis{oValveMoveToEndPos} become \true, or the system does not become disabled, \inlinecordis{oEnabled} becomes \false.

\item  \label{req:if-MOVE_TO_ZERO_POSITION-inevitably-oInZeroPosition-or-oValveMoveToZeroPos}
Whenever the system is enabled and command \inlinecordis{MOVE_TO_ZERO_POSITION} is accepted, inevitably either output signal \inlinecordis{oInZeroPosition} becomes \true\ or output \inlinecordis{oValveMoveToZeroPos} becomes \true, or the system becomes disabled, unless another command is accepted or rejected.

This requirement and its formalization are completely symmetric to requirement~\ref{req:if-MOVE_TO_END_POSITION-inevitably-oInEndPosition-or-oValveMoveToEndPos}. Note that, accepting command \inlinecordis{MOVE_TO_ZERO_POSITION} is encoded as \inlinemcrl{command(6,true)}. The property is then formalized as follows.
\begin{lstlisting}[language=mcrl2]
nu Y(enabled: Bool = false) .
  (forall b: Bool . <state_M2'oEnabled(b)>true
                                                                => [!command(6,true)]Y(b)) &&
  ([exists b: Bool . state_M2'oEnabled(b)]false
                                                                => [!command(6,true)]Y(enabled)) &&
  (val(enabled) && <state_M2'oEnabled(false)>true
                                                                => [command(6,true)]Y(false)) &&
  (val(enabled) && [state_M2'oEnabled(false)]false
                                                                => [command(6,true)]
    mu X.
      [!(
    			(exists i: Nat, b:Bool . command(i, b)) ||
				(exists b: Bool . state_M2'oValveMoveToZeroPos(b) ||
										            state_M2'oValveMoveToEndPos(b) ||
										            state_M2'oInZeroPosition(b) ||
										            state_M2'oInEndPosition(b) ||
  										            state_M2'oEnabled(b)) ||
  				(exists i: Nat, l: List(Nat) . states(i, l))
      )]X ||
      <state_M2'oInZeroPosition(true) ||
            state_M2'oValveMoveToZeroPos(true)>true ||
      <state_M2'oEnabled(false)>true
    )
\end{lstlisting}

\item \label{req:if-oValveMoveToEndPos-then-oInEndPosition}
Whenever the system is enabled, and output \inlinecordis{oValveMoveToEndPos} becomes \true, and as long as the system stays enabled, it is possible to reach the end position (i.e., \inlinecordis{oInEndPosition} becomes \true), unless another command is written on the interface.

This requirement is similar in nature to requirement~\ref{req:if-MOVE_TO_END_POSITION-inevitably-oInEndPosition-or-oValveMoveToEndPos}.
However, its formulation is much weaker. Ideally, we may expect that stronger requirement that inevitably \inlinecordis{oInEndPosition} is reached also holds.
Yet, in order to reach that situation, the system depends on the values of the input and input signals. In particular, it requires that eventually \inlinecordis{iEndPosSensor} becomes \true.
We have therefore weakened the requirement to say that \inlinecordis{oInEndPosition} can become true, but that this is not required.

This is formalized as follows.

\begin{lstlisting}[language=mcrl2]
[true*](<state_M2'oValveMoveToEndPos(false)>true =>
  [true*](([state_M2'oValveMoveToEndPos(true)]false
                        && <state_M2'oEnabled(true)>true) =>
    [true*](([state_M2'oEnabled(false)]false
                        && <state_M2'oValveMoveToEndPos(true)>true) =>
      ([!(exists i: Nat. freecmd(i))*]
        <!(exists i: Nat. freecmd(i))*.state_M2'oInEndPosition(true)>true))))
\end{lstlisting}

The formula can be explained as follows.
The left-hand side of the first implication, \inlinemcrl{[true*](<state_M2'oValveMoveToEndPos(false)>true}, indicates that for all sequences of zero or more actions, if \inlinemcrl{oValveMoveToEndPos} is eventually set to \false, then the second implication should hold.
The second implication is similar, stating that for all sequences of zero or more actions, if there is no transition indicating that \inlinemcrl{oValveMoveToEndPos} is \true\ and if \inlinemcrl{oEnabled} is set to \true, then the third implication should hold.
The third implication is, again, similar.
If all three conditions have been met (in the appropriate order), then along every execution on which no new command has been issued to the cylinder, it is possible to reach the end position. This is expressed by the last part of the formula.

\item \label{req:if-oValveMoveToZeroPos-then-oInZeroPosition}
Whenever the system is enabled, and  \inlinemcrl{oValveMoveToZeroPos} becomes \true, and as long as the system stays enabled,
it is possible to reach the zero position (i.e., \inlinemcrl{oInZeroPosition} becomes \true), unless another command is written on the interface.
This requirement and its formalization are completely symmetric to requirement~\ref{req:if-oValveMoveToEndPos-then-oInEndPosition} and is formalized as follows.
\begin{lstlisting}[language=mcrl2]
[true*](<state_M2'oValveMoveToZeroPos(false)>true =>
  [true*](([state_M2'oValveMoveToZeroPos(true)]false
                        && <state_M2'oEnabled(true)>true) =>
    [true*](([state_M2'oEnabled(false)]false
                        && <state_M2'oValveMoveToZeroPos(true)>true) =>
      ([!(exists i: Nat. freecmd(i))*]
        <!(exists i: Nat. freecmd(i))*.state_M2'oInZeroPosition(true)>true))))

\end{lstlisting}

\item \label{req:if-not-iEndPosSensor-and-not-iForceToEnd-and-not-MOVE_TO_END_POS-or-not-TOGGLE-then-not-oInEndPos}
When the system initializes, the state \inlinecordis{InEndPosition} should not be reached spontaneously. We therefore require that output signal \inlinecordis{oInEndPosition} remains \false, as long as both \inlinecordis{iForceEnablingToEndPos} is \false\ and either \inlinecordis{iEndPosSensorConnected} or \inlinecordis{iEndPosSensor} is \false, and commands \inlinemcrl{MOVE_TO_END_POSITION} and \inlinemcrl{TOGGLE} have not been accepted.
This is formalized as follows.
\begin{lstlisting}[language=mcrl2]
nu X(iEndPosSensor: Bool = false, iEndPosSensorConnected: Bool = false,
		 iForceEnablingToEndPos: Bool = false) .
  (forall iEndPosSensor': Bool .
		[exists iZPS: Bool . inputs(iZPS, iEndPosSensor')]
			X(iEndPosSensor', iEndPosSensorConnected, iForceEnablingToEndPos)) &&
	(forall iForceEnablingToEndPos', iEndPosSensorConnected': Bool .
		[exists iCAOK, iZPSC, iROOES, iFETZP: Bool .
			free_input_signals(iCAOK, iZPSC, iEndPosSensorConnected',
												 iROOES, iFETZP, iForceEnablingToEndPos')]
				X(iEndPosSensor, iEndPosSensorConnected', iForceEnablingToEndPos')) &&
		(val(iForceEnablingToEndPos ||
					(iEndPosSensorConnected && iEndPosSensor)) ||
			([!((exists iZS, iES: Bool . inputs(iZS, iES)) ||
				(exists iC, iZSC, iESC, iR, iFZ, iFEP: Bool .
					free_input_signals(iC, iZSC, iESC, iR, iFZ, iFEP)) ||
				command(7, true) ||
				command(8, true))
			 ]X(iEndPosSensor, iEndPosSensorConnected, iForceEnablingToEndPos) &&
			 [state_M2'oInEndPosition(true)]false))
\end{lstlisting}
Note that in the translation, \inlinemcrl{command(7,true)} and \inlinemcrl{command(8,true)} encode the acceptance of commands \inlinecordis{MOVE_TO_END_POSITION} and \inlinecordis{TOGGLE}, respectively.
As the condition that releases the obligation for \inlinecordis{oInEndPosition} to be \false\ refers to both inputs and input signals, the invariant formula keeps track of the most recent value these particular inputs and input signals were assigned.
The first conjunct tracks this by, after every action \inlinemcrl{inputs}, updating parameter \inlinemcrl{iEndPosSensor} with the value set in the action. The second conjunct does the same for the input signals.
The third conjunct is the key to checking the actual requirement.
It first checks whether the most recent values are such that the requirement has been fulfilled, i.e., either \inlinecordis{iForceEnablingToEndPos} has become \true, or \inlinecordis{iEndPosSensorConnected} and \inlinecordis{iEndPosSensor} are both \true. If this is not the case, we need to ensure that \inlinecordis{oInEndPosition} is not \true, which is checked using \inlinemcrl{[state_M2'oInEndPosition(true)]false}, and that for all executions other than those accepting commands \inlinecordis{MOVE_TO_END_POSITION} and \inlinecordis{TOGGLE}, recursively the requirement again holds.

\item \label{req:if-not-iZeroPosSensor-and-not-iForceToZero-and-not-MOVE_TO_ZERO_POS-or-not-TOGGLE-then-not-oInZeroPos}
When the system initializes, the state \inlinecordis{InZeroPosition} should not be reached spontaneously. We therefore require that output signal \inlinecordis{oInZeroPosition} remains \false, as long as both \inlinecordis{iForceEnablingToZeroPos} is \false\ and either \inlinecordis{iZeroPosSensorConnected} or \inlinecordis{iZeroPosSensor} is \false, and commands \inlinemcrl{MOVE_TO_ZERO_POSITION} and \inlinemcrl{TOGGLE} have not been accepted.

This requirement and its formalization are completely symmetric to requirement~\ref{req:if-not-iEndPosSensor-and-not-iForceToEnd-and-not-MOVE_TO_END_POS-or-not-TOGGLE-then-not-oInEndPos}.
It is formalized as follows.

\begin{lstlisting}[language=mcrl2]
nu X(iZeroPosSensor: Bool = false, iZeroPosSensorConnected: Bool = false,
		 iForceEnablingToZeroPos: Bool = false) .
  (forall iZeroPosSensor': Bool .
		[exists iEPS: Bool . inputs(iZeroPosSensor', iEPS)]
			X(iZeroPosSensor', iZeroPosSensorConnected, iForceEnablingToZeroPos)) &&
	(forall iForceEnablingToZeroPos', iZeroPosSensorConnected': Bool .
		[exists iCAOK, iEPSC, iROOES, iFETEP: Bool .
			free_input_signals(iCAOK, iZeroPosSensorConnected', iEPSC,
												 iROOES, iForceEnablingToZeroPos', iFETEP)]
				X(iZeroPosSensor, iZeroPosSensorConnected',
				                                 iForceEnablingToZeroPos')) &&
		(val(iForceEnablingToZeroPos ||
					(iZeroPosSensorConnected && iZeroPosSensor)) ||
			([!((exists iZS, iES: Bool . inputs(iZS, iES)) ||
				(exists iC, iZSC, iESC, iR, iFZ, iFEP: Bool .
					free_input_signals(iC, iZSC, iESC, iR, iFZ, iFEP)) ||
				command(6, true) ||
				command(8, true))
			 ]X(iZeroPosSensor, iZeroPosSensorConnected, iForceEnablingToZeroPos) &&
			 [state_M2'oInZeroPosition(true)]false))
\end{lstlisting}

\item \label{req:if-oInEndPosition-and-iEndPosSensor-and-iZeroPosSensor-then-not-oInEndPosition}
Whenever output signal \inlinemcrl{oInEndPosition} is \true, and input \inlinemcrl{iEndPosSensor} and input \inlinemcrl{iZeroPosSensor} are both \true, inevitably \inlinemcrl{oInEndPosition} becomes \false\ unless input \inlinemcrl{iZeroPosSensor} becomes \false\ or input signal \inlinemcrl{iZeroPosSensorConnected} becomes \false.

This is formalized as follows.
\begin{lstlisting}[language=mcrl2]
nu X(inEndPos: Bool = false, iZeroPos: Bool = false, iEndPos: Bool = false) .
    (forall e: Bool . <state_M2'oInEndPosition(e)>true =>
        (forall b,c: Bool . [inputs(b, c)]X(e,b,c) &&
        [!exists a1,a2: Bool .inputs(a1, a2)]X(e,iZeroPos,iEndPos))) &&
    (forall e: Bool . [state_M2'oInEndPosition(e)]false =>
        (forall b,c: Bool . [inputs(b, c)]X(inEndPos,b,c) &&
        [!exists a1,a2: Bool .inputs(a1, a2)]X(inEndPos,iZeroPos,iEndPos))) &&
    (val(inEndPos && iZeroPos && iEndPos) =>
        mu X.[!((exists iESP:Bool . inputs(false,iESP)) ||
                    (exists a,c,d,e,f:Bool . free_input_signals(a,false,c,d,e,f)) ||
                    (exists b:Bool . state_M2'oValveMoveToZeroPos(b) ||
                                                   state_M2'oValveMoveToEndPos(b) ||
                                                   state_M2'oInZeroPosition(b) ||
                                                   state_M2'oInEndPosition(b) ||
                                                   state_M2'oEnabled(b)) ||
                    (exists i:Nat,l:List(Nat) . states(i,l))
        )]X || <state_M2'oInEndPosition(false)>true)
\end{lstlisting}

This property is similar to requirements~\ref{req:if-not-oEnabled-and-iCompressedAirOk-then-oEnabled} and~\ref{req:if-MOVE_TO_END_POSITION-inevitably-oInEndPosition-or-oValveMoveToEndPos}.
The formula uses a greatest fixed point (nu) and a least fixed point (mu). The greatest fixed point is parameterized by the Boolean variables \inlinemcrl{inEndPos}, \inlinemcrl{iZeroPos} and \inlinemcrl{iEndPos} that are used to keep track of whether output signal \inlinemcrl{oInEndPosition} or input \inlinemcrl{iEndPosSensor} or input \inlinemcrl{iZeroPosSensor} have become \true, respectively.
In order to keep track of these values, we distinguish two cases as in Requirement~\ref{req:if-not-oEnabled-and-iCompressedAirOk-then-oEnabled}.
Now, if \inlinemcrl{oInEndPosition} is \true, and both \inlinemcrl{iEndPosSensor} and \inlinemcrl{iZeroPosSensor} are \true, the least fixed point subformula needs to hold.
In the least fixed point subformula we need to exclude the paths through the self-loops and those in which input \inlinemcrl{iZeroPosSensor} or input signal \inlinemcrl{iZeroPosSensorConnected} becomes \false.

\item \label{req:not-both-valves-open-simultaneously}
Invariantly, \inlinecordis{oValveMoveToZeroPos} and \inlinecordis{oValveMoveToEndPos} are not \true\ at the same time.

This formalized as follows.
\begin{lstlisting}[language=mcrl2]
[true*]!(<state_M2'oValveMoveToZeroPos(true)>true &&
    <state_M2'oValveMoveToEndPos(true)>true)
\end{lstlisting}

This is a straightforward invariant that relies on the self-loops exposing the values of the outputs.
\end{enumerate}

\section{Additional state machine diagrams}\label{sec:extra-diagrams}
\begin{figure}[!h]
\centering
\includegraphics[width=0.88 \linewidth]{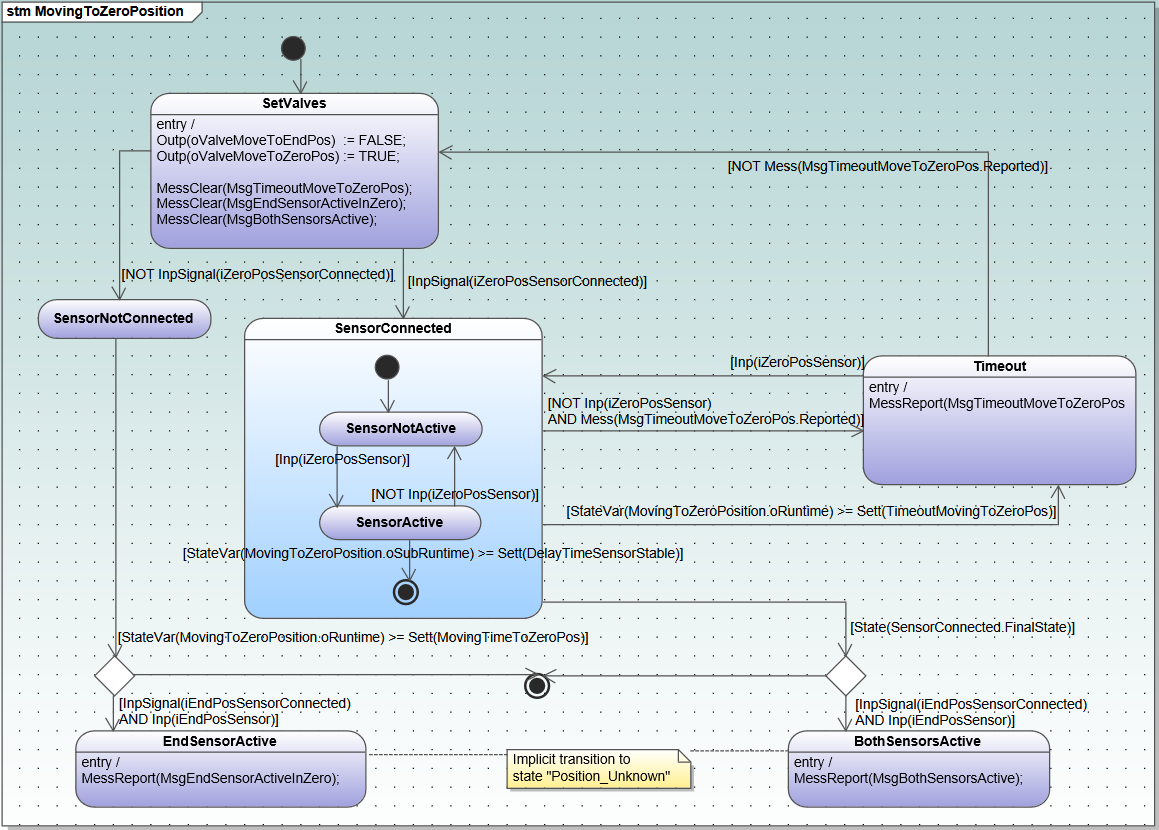}
\caption{Substatemachine \inlinecordis{MovingToZeroPosition}}
\label{fig:cylinder-movingtozeroposition-sm}
\end{figure}

\begin{figure}[!h]
\centering
\includegraphics[width=0.9\linewidth]{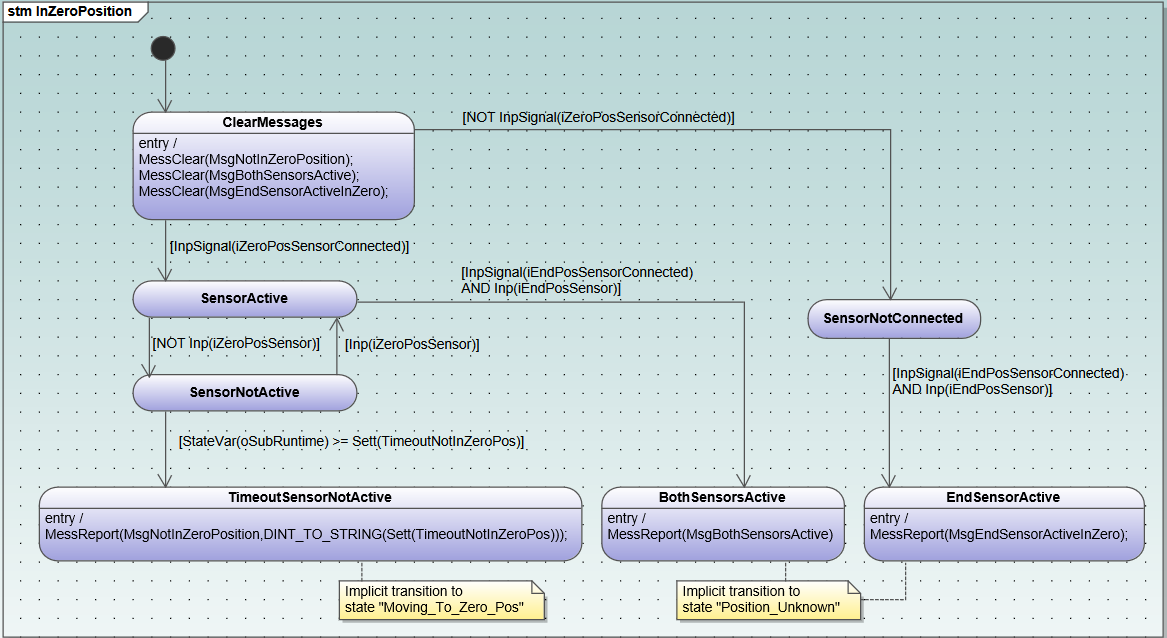}
\caption{State machine diagram \inlinecordis{InZeroPosition}}
\label{fig:cylinder-inzeroposition-sm}
\end{figure}

For the sake of completeness, we here include the remaining two state machine diagrams \sm{MovingToZeroPosition}, in Figure~\ref{fig:cylinder-movingtozeroposition-sm}, and \sm{InZeroPosition}, in Figure~\ref{fig:cylinder-inzeroposition-sm}.

Substatemachine \sm{MovingToZeroPosition} is symmetric to substatemachine \sm{MovingToEndPosition}, in Figure~\ref{fig:cylinder-movingtoendposition-sm}. Subdiagram \sm{InZeroPosition} is symmetric to \sm{InEndPosition}, which was shown in Figure~\ref{fig:cylinder-inendposition-sm}. We refer to Section~\ref{sec:cylinder} for an explanation of the state machine diagrams.
\vfill

\end{document}